\documentclass[aps,prd,reprint]{revtex4-1}

\usepackage{lineno}
\usepackage{subfigure}
\usepackage{adjustbox}
\usepackage{amsmath}  
\usepackage{amsfonts} 
\usepackage{graphicx} 
\usepackage{multirow}
\usepackage{gensymb}
\usepackage{textcomp}
\usepackage{color}
\usepackage{setspace}

\usepackage{floatrow}
\usepackage{feynmp}
\RequirePackage[colorlinks,citecolor=blue,urlcolor=blue,linkcolor=blue]{hyperref}
\RequirePackage{url}

\begin{document}

\title{Sensitivity on Anomalous Neutral Triple Gauge Couplings via  $ZZ$  Production at FCC-hh}

\author{A.~Yilmaz} \email{aliyilmaz@giresun.edu.tr}
\affiliation{Department of Electrical and Electronics Engineering, Giresun University, 28200, Giresun, Turkey}
\author{A.~Senol} \email{senol\_a@ibu.edu.tr}
\author{H.~Denizli}  \email{denizli\_h@ibu.edu.tr}
\affiliation{Department of Physics, Bolu Abant Izzet Baysal University, 14280 Bolu, Turkey}
\author{I.~Turk Cakir} \email{ilkay.turk.cakir@cern.ch}
\affiliation{Department of Energy Systems Engineering, Giresun University, 28200 Giresun, Turkey}
\author{O.~Cakir} \email{ocakir@science.ankara.edu.tr}
\affiliation{Department of Physics, Ankara University, 06100 Ankara, Turkey}

\begin{abstract}

We study the sensitivity of anomalous neutral triple gauge couplings ($aNTGC$) via  $pp \rightarrow ZZ$  production in the 4$\ell$ channel at 100 TeV centre of mass energy of future circular hadron collider,  \verb"FCC-hh". 
The analysis including the realistic detector effects is performed in the mode where both Z bosons decay into same flavor, oppositely charged lepton pairs. The sensitivities to the charge-parity (CP)-conserving $C_{\tilde{B}W} / \Lambda^{4}$ and CP-violating $C_{WW}  / \Lambda^{4}$, $C_{BW} / \Lambda^{4}$ and $C_{BB} / \Lambda^{4}$ couplings obtained at 95\% Confidence Level (C.L.) using the invariant mass distribution of 4$\ell$ system reconstructing the leading and sub-leading Z boson candidates are $[-0.117, \,\, +0.117]$,  $[-0.293, \,\, +0.292]$, $[-0.380, \,\, +0.379]$,  and $[-0.138, \,\, +0.138]$ in the unit of TeV$^{-4}$, respectively.

\end{abstract}
\maketitle

\section{Introduction}
     
The studies on the diboson production at colliders play an important role in testing  the non-Abelian SU (2)$_{L} \times U (1)_{Y}$ gauge group of the electroweak sector in the Standard Model (SM) and searching for new phenomena at the TeV-energy scale~\cite{Neubauer_2011}.
Since  there is no triple gauge couplings between the photon and $Z$ boson  ($Z\gamma \gamma$ and $Z\gamma Z$) except $WWZ$ and $WW\gamma$  in the SM, pairs of $Z$ bosons cannot be created at a single vertex in the SM. 
Therefore any deviations from SM predictions on neutral triple gauge couplings (including $ZZ\gamma$,  $Z\gamma \gamma$ and $ZZZ$ vertices) can give an indication about new physics beyond the SM.
 The new physics effects at high energy can be parametrized in the Effective Field Theory (EFT) approach. This theory is general enough to point  the most probable places to observe these effects since it is renormalizable, includes the gauge symmetries of the standard model and can be used at both tree level and loop level. There is no concern on violating of the unitary  of anomalous couplings  in scattering processes at higher energies according to this theory.
 Anomalous NTG vertices can be added in an effective Lagrangian using EFT approach and parametrized by  CP-conserving and CP-violating  couplings, while  no SM NTGC is present at tree-level~\cite{Green:2016trm}. 
 
The production of $ZZ$ dibosons in the 4$\ell$ final state  have been studied by  various collaborations  such as the Large Electron-Positron (LEP)~\cite{Barate:1999jj, Acciarri:1999ug,  Abdallah:2003dv, Abbiendi:2003va, Alcaraz:2006mx, Schael:2013ita} where the first bounds on anomalous neutral triple gauge couplings (aNTGCs)  using e$^+$e$^-$ collider was obtained, the Collider Detector at Fermilab (CDF)~\cite{CDF:2011ab, PhysRevD.89.112001} and $D\O$~\cite{Abazov:2007ad, D0:2013rca} also searched the limits of aNTGC at Tevatron $pp$ collider.  Recently, ATLAS~\cite{Aad:2015zqe, PhysRevD.97.032005} and CMS~\cite{Aaboud:2017rwm, Sirunyan_2018} collaborations published the improved limits of aNTGCs thanks to the center of mass energy of LHC in the range of 13 TeV at the LHC. This high center of mass energy leads to enhance the cross-section which would widen the range of triple gauge coupling studies. There are also some phenomenological studies  for probing the  sensitivities of aNTGCs at hadron colliders in the EFT framework~\cite{Senol_2014, Mangano_2016, Frye_2016, DORIGO2018211, SENOL2018365}.

 The dimension-eight (dim-8) effective  Lagrangian for nTGC in the scope of EFT assuming the local U(1)$_{EM}$ and Lorentz symmetry  can be written as~\cite{Degrande_2014}

\begin{equation}
\mathcal{L}^{nTGC} = \mathcal{L}_{SM} + \sum_{i} \frac{C_{i}}{\Lambda^{4}} (\mathcal{O}_{i} + \mathcal{O}_{i}^{\dagger})
\label{eqn:lagrangian}
\end{equation}

where $i$ is the index of equations running over the operators given as
\begin{eqnarray}
\mathcal{O}_{\widetilde{B}W}  & = & i H^{\dagger} \widetilde{B}_{\mu\nu} W^{\nu \rho} \{D_{\rho}, D^{\nu} \} H, \\
\mathcal{O}_{BW}  & = & i H^{\dagger} B_{\mu\nu}W^{\nu \rho} \{D_{\rho}, D^{\nu} \} H, \\
\mathcal{O}_{WW}  & = & i H^{\dagger} W_{\mu\nu} W^{\nu \rho} \{D_{\rho}, D^{\nu} \} H, \\
\mathcal{O}_{BB}  & = & i H^{\dagger} B_{\mu\nu} B^{\nu \rho} \{D_{\rho}, D^{\nu} \} H.
\label{eqn:lagrangian2}
\end{eqnarray}

where $\widetilde{B}_{\nu\mu}$ is dual $B$ strength tensor. We used the convention given below in the definitions of the operators
\begin{eqnarray}
B_{\mu\nu} & = & (\partial_{\mu} B_{\nu} - \partial_{\nu} B_{\mu}) \\
W_{\mu\nu} & = & \sigma^{I} (\partial_{\mu} W^{I}_{\nu} - \partial_{\nu} W^{I}_{\mu} + g\epsilon_{IJK} W^{J}_{\mu} W^{K}_{\nu} )
\label{eqn:lagrangian3}
\end{eqnarray}

with $\langle \sigma^{I} \sigma^{J} \rangle = \partial^{I \, J} / 2$  and

\begin{equation}
D_{\mu} \equiv \partial_{\mu} - i g_{w} W^{i}_{\mu} \sigma^{i} - i \frac{g^{\prime}}{2} B_{\mu} Y 
\label{eqn:lagrangian4}
\end{equation}

The coefficients of these four dimension-eight operators describing  aNTGC are CP-conserving  $C_{\tilde{B}W} / \Lambda^{4}$ and  CP-violating  $C_{WW}  / \Lambda^{4}$, $C_{BW} /   \Lambda^{4}$ and $C_{BB}  /  \Lambda^{4}$ couplings.  

The current limits  on $C_{\tilde{B}W} / \Lambda^{4}$, $C_{WW}  / \Lambda^{4}$, $C_{BW} /   \Lambda^{4}$ and $C_{BB}  /   \Lambda^{4}$ couplings of   \texttt{dim-8} operators  converted from the couplings of  \texttt{dim-6} operators for the process $pp \rightarrow ZZ \rightarrow  \ell^+ \ell^- \ell^{\prime +} \ell^{\prime -} $~\cite{PhysRevD.97.032005}  where $\ell = e$  or  $\mu$ and  $Z\gamma \rightarrow \nu \bar{\nu}\gamma$~\cite{ATLAS:2018eke}  at the center of mass energy $\sqrt{s} = 13$ TeV and integrated luminosity $L_{int} = 36.1$ fb$^{-1} $  from the LHC are given in Table~\ref{tab:limits1}. In this table, all couplings other than the one under study are set to zero.

\begin{table}[h!]
\centering 
\begin{tabular}{  lccc  }
 \hline 
 Couplings &  \multicolumn{3}{c}{Limit 95\% C.L.}   \\
  $ (TeV^{-4})$   & &  $ZZ \rightarrow 4 \ell$~\cite{PhysRevD.97.032005} & $Z\gamma \rightarrow \nu \bar{\nu}\gamma$~\cite{ATLAS:2018eke} \\
\hline
$C_{\tilde{B}W}  / \Lambda^{4}$ & & $-5.9, +5.9$ & $-1.1, \,\, +1.1$ \\
$C_{WW} / \Lambda^{4}$ & & $-3.0, +3.0$ & $-2.3, \,\, +2.3$ \\
$C_{BW} / \Lambda^{4}$ & & $-3.3, +3.3$ &  $-0.65, +0.64$\\
$C_{BB} / \Lambda^{4}$ & & $-2.7, +2.8$ & $-0.24, +0.24$ \\
\hline 
\end{tabular} 
\caption{Observed one dimensional 95\% C.L. limits on $C_{\tilde{B}W} / \Lambda^{4}$, $C_{WW}  / \Lambda^{4}$, $C_{BW} /   \Lambda^{4}$ and $C_{BB}  /   \Lambda^{4}$ EFT parameters from LHC.}
\label{tab:limits1}
\end{table}

The future circular collider project, FCC~\cite{FCCweb}, proposed to have three collider options (FCC-ee, FCC-eh and FCC-hh) working at different center of mass energies. The hadron collider option of FCC (FCC-hh) is planned to reach an integrated luminosity of 20-30 ab$^{-1}$ at 100 TeV center mass energy. 
FCC-hh, comparing to LHC, has the energy scale by a factor about 7 depending on the process~\cite{Benedikt:2651300}. 

Exploring the new physics effects in the production of diboson is a challenging task.
In the literature ZZ diboson production has been examined in 2 decay channels such as ``2$\ell$2$\nu$" and ``4$\ell$" channel~\cite{Green:2016trm}. In the first channel $ZZ \rightarrow$ 2$\ell$2$\nu$,  one of the Z decays into a neutrino while  the other one decaying into a same flavor, oppositely-charged two leptons which leads to increase in the missing transfer energy in the final state. Therefore this channel exposes to a larger background contribution and it is not kinematically reconstructable completely. In the second decay channel $ZZ \rightarrow$ 4$\ell$, not only the first Z boson, but also the other Z boson decays into a same-flavor, oppositely charged two leptons. This process gives rise to include a very low background and kinematically reconstructable in the final state. On the other hand, one needs to take into account the process has small branching fractions results with a low statistics in the final state.

This paper will be organized as follows: 
In section~\ref{sec:simulation} we will discuss the simulation environment of $ZZ$ diboson production for signal and background at FCC-hh collider. Event selection procedures of our phenomenological study in the 4$\ell$ final state will be given in section~\ref{sec:eventSelect}. 
In section~\ref{sec:results}, we will give the collected results for  4$\ell$ final state analysis.
Conclusions on the sensitivities  of each couplings will be summarized in section~\ref{sec:conclusion}.

\section{Generation of signal and background events} \label{sec:simulation}

To obtain the bounds on aNTGC parameters of $ZZ$ diboson production  in the framework of the EFT at the FCC-hh. We generated signal and background events for the $pp \rightarrow ZZ $ process by importing the  signal \verb"aTGC" implemented  through \verb"UFO" model file into \verb"MadGraph5_aMC@NLO v2.6.4"~\cite{Alwall_2014}. The \verb"PYTHIA v8.2"~\cite{SJOSTRAND2015159} package is used for parton showering and hadronization.
 \verb"LHAPDF v6.1.6"~\cite{Buckley_2015} library and its \verb"NNPDF v2.3"~\cite{BALL2013244} set is used as the default set of parton distribution functions (PDFs) for all simulated MC samples.  $3 \times 10^{6}$ events of the signal and the background were generated for each \texttt{dim-8} couplings. The detector response is simulated using a detailed description of the FCC-hh detector card implemented  in the \verb"Delphes v3.4.1"~\cite{de_Favereau_2014}. All events are analyzed by using the \verb"ExRootAnalysis"~\cite{exRootAnalysis} package  with \verb"ROOT v6.16"~\cite{BRUN199781}. The kinematical distributions are normalized to the number of expected events which is defined to be the cross section of each processes including the branchings times integrated luminosity of $\mathcal{L}_{int}$ = 10 ab$^{-1}$.

\begin{figure}[!htb] 
 	\centering 
	\subfigure[signal]{%
 		  \includegraphics[width=0.48\textwidth]{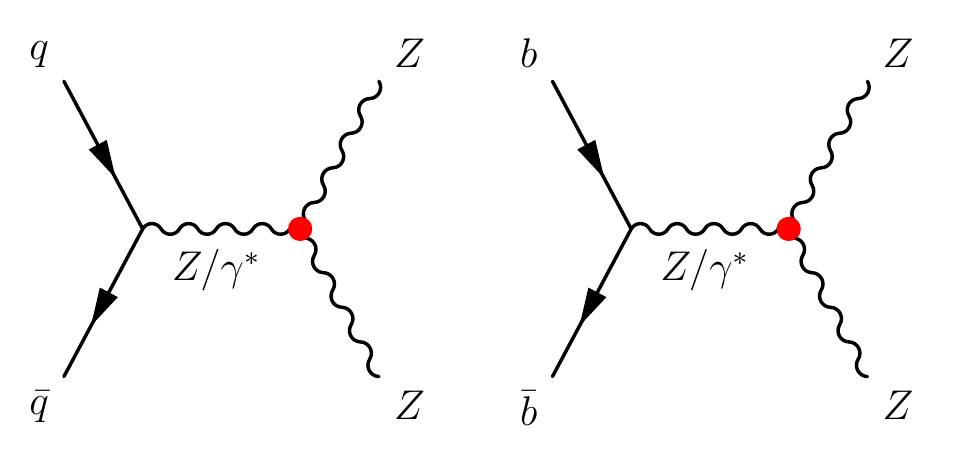}
		\label{fig:sig}
    	}
    	\subfigure[bakground]{%
   		   \includegraphics[width=0.47\textwidth]{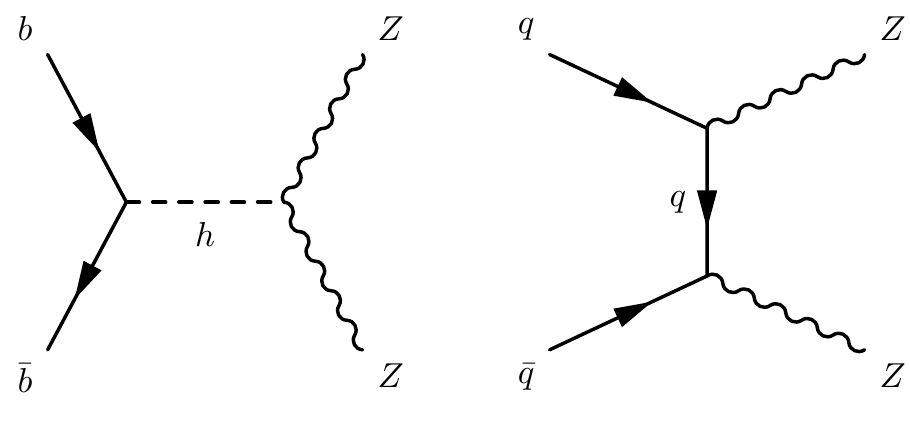}
		\label{fig:bck}
    	}
  \caption{Feynmann diagrams of ZZ production  (a) for signal including an aNTGC vertex depicted by a red dot, (b) for the SM background.} 
   \label{fig:feynmanDiag}
\end{figure}

Feynman diagrams that contribute to the signal and its main-background processes are shown in Fig.~\ref{fig:sig} and Fig.~\ref{fig:bck}, respectively. The red dot represents the aNTGC vertex in the production of $ZZ$.

The cross section is calculated with a set of generator level cuts;
a lepton is declared to be isolated if the $p_{T}$-sum of all particles within the isolation cone size R$_{iso}$ = 0.3, minimum $p_{T}$ = 10 GeV  and $|\eta| <$ 2.5 for the charged leptons. In the calculations, default mass of the $Z$ boson is used as 91.187 GeV.

\begin{figure} [hbt]

 \centering
 
   \includegraphics[width=0.8\textwidth]{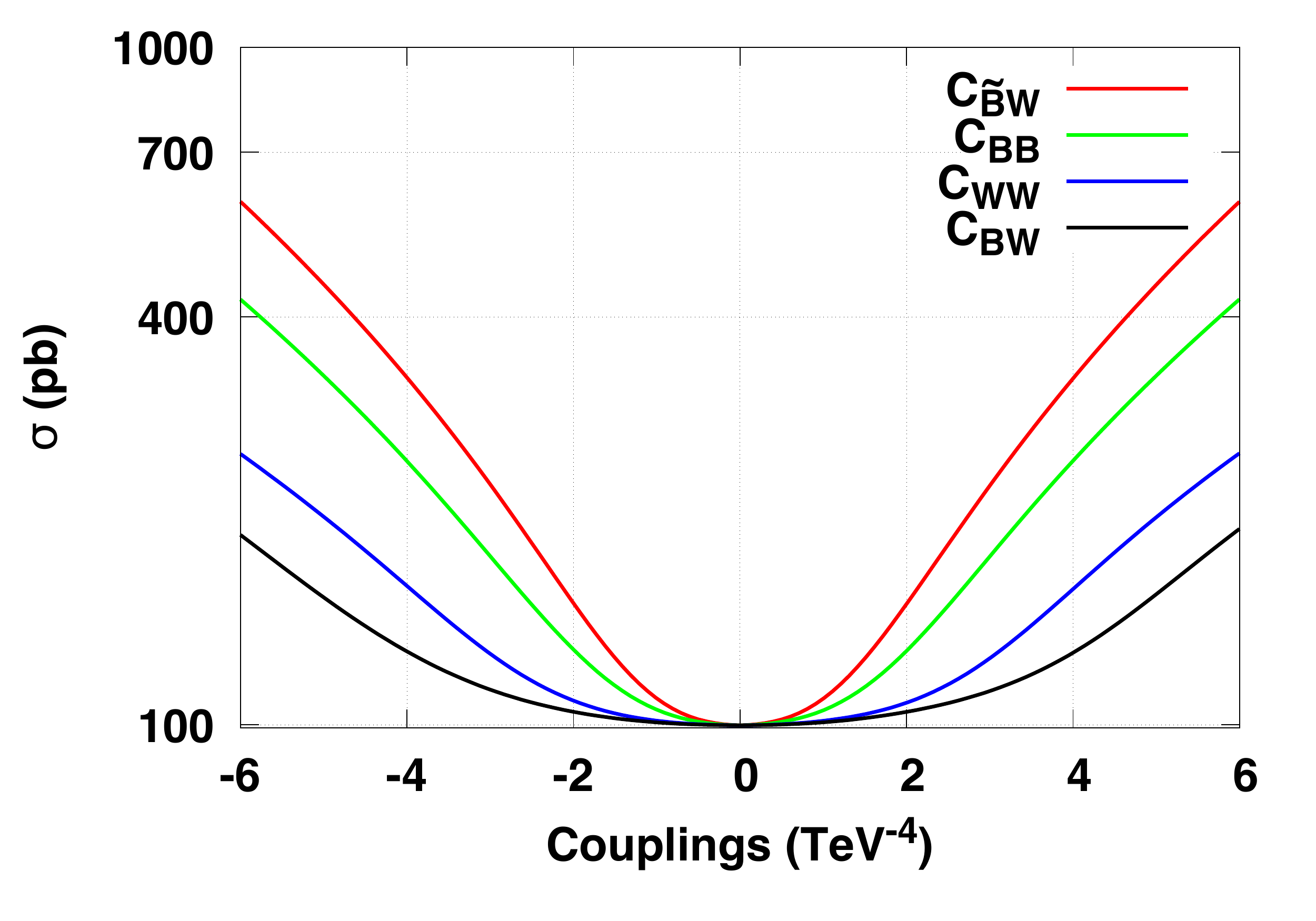}
  \caption{Cross sections for the process with aNTGCs including CP-conserving and CP-violating terms in the Lagrangian.} 
   \label{fig:xSectionPlot}
\end{figure}

The cross sections of the $ZZ$ process as a function of mentioned four \texttt{dim-8} couplings are shown in Fig.~\ref{fig:xSectionPlot}. In this figure, only one coupling at a time is varied from its SM value and plotted as a function of couplings in the range of limits reported by CMS Collaboration~\cite{Sirunyan_2018}.  One can clearly see the deviation from the SM.

\section{Event selection} \label{sec:eventSelect}

We consider 4$\ell$ final state in our analysis based on Ref.~\cite{Sirunyan_2018} including three possible options; $e^{+}e^{-}e^{+}e^{-}$, $\mu^+\mu^-\mu^+\mu^-$, and $e^+e^-\mu^+\mu^-$.  
The preselection for this analysis require the presence of a pair of leptons of the same or different flavors~\cite{Khachatryan_2017}.  
 All permutations of leptons giving a pair of $Z / \gamma^\ast$  candidates are considered  within each event. The pairing ambiguity is resolved by ordering the pair of dilepton candidates  based on the differences between the reconstructed invariant mass  of dilepton canditate ($m_{\ell^+ \ell^-}$) and nominal $Z$ boson  mass $m_{Z}$. Therefore, the dilepton candidate with an invariant mass closest to the nominal $Z$ boson mass~\cite{PhysRevD.98.030001}, is denoted $leading \, Z$ while the second closest  is defined as $subleading \, Z$. 

\begin{figure}[!htb] 
 	\centering 
	\subfigure[ ]{
    		\includegraphics[width=0.48\textwidth]{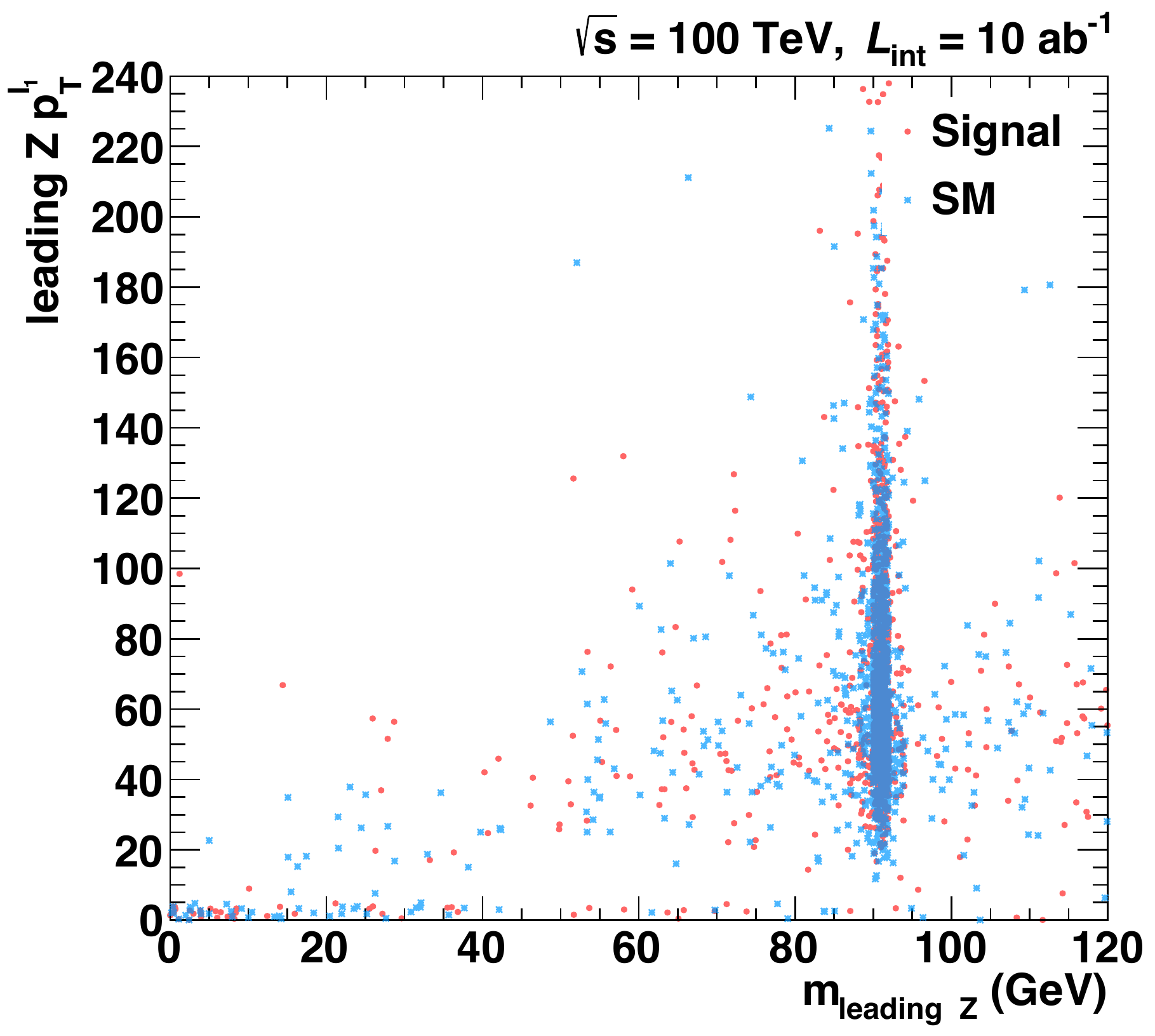}
		\label{fig:e_leadingZmassPTl1cut0}
    	}
    	\subfigure[]{
    		\includegraphics[width=0.48\textwidth]{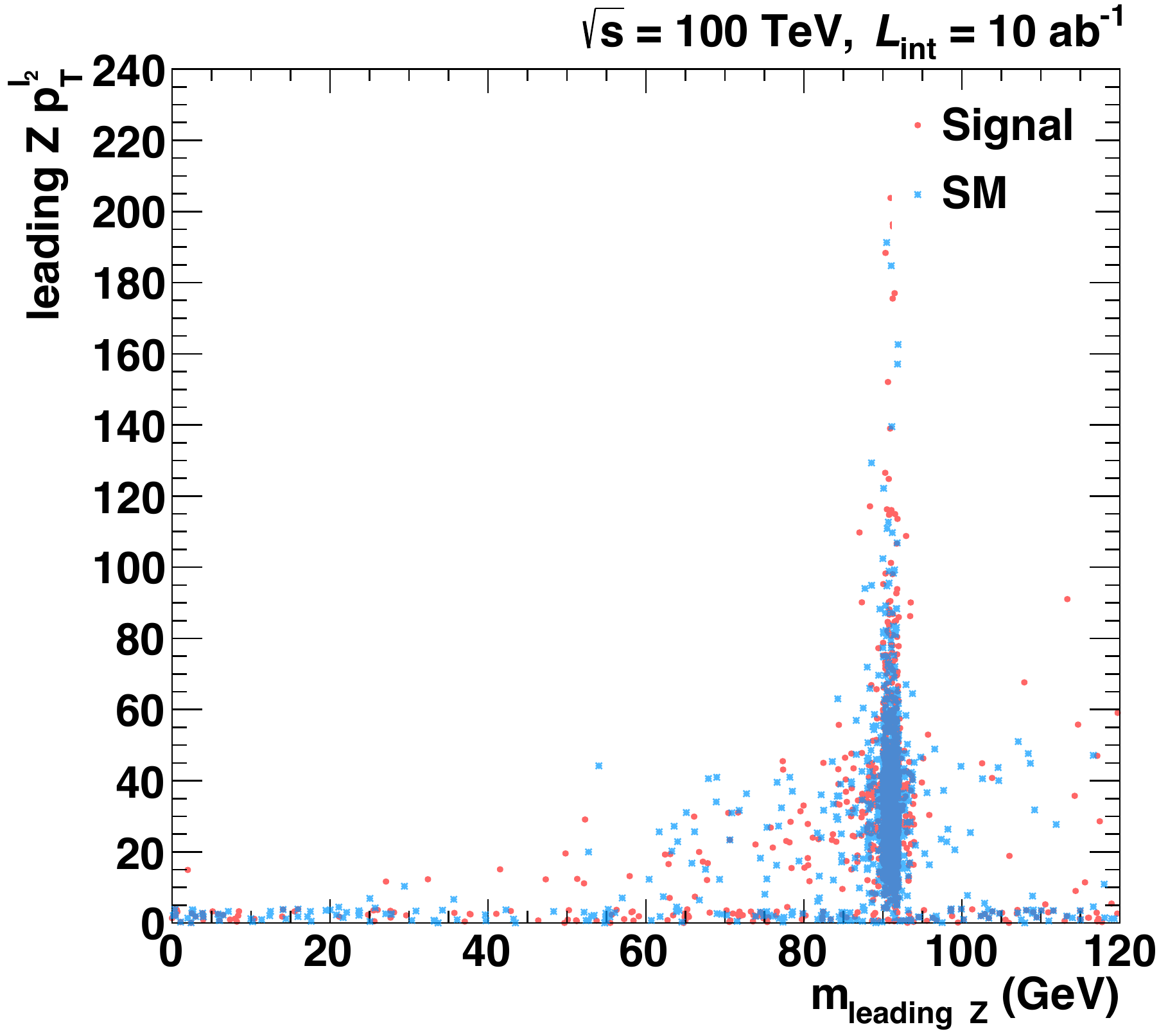}
		\label{fig:e_leadingZmassPTl2cut0}
    	}
    	\subfigure[]{
    		\includegraphics[width=0.48\textwidth]{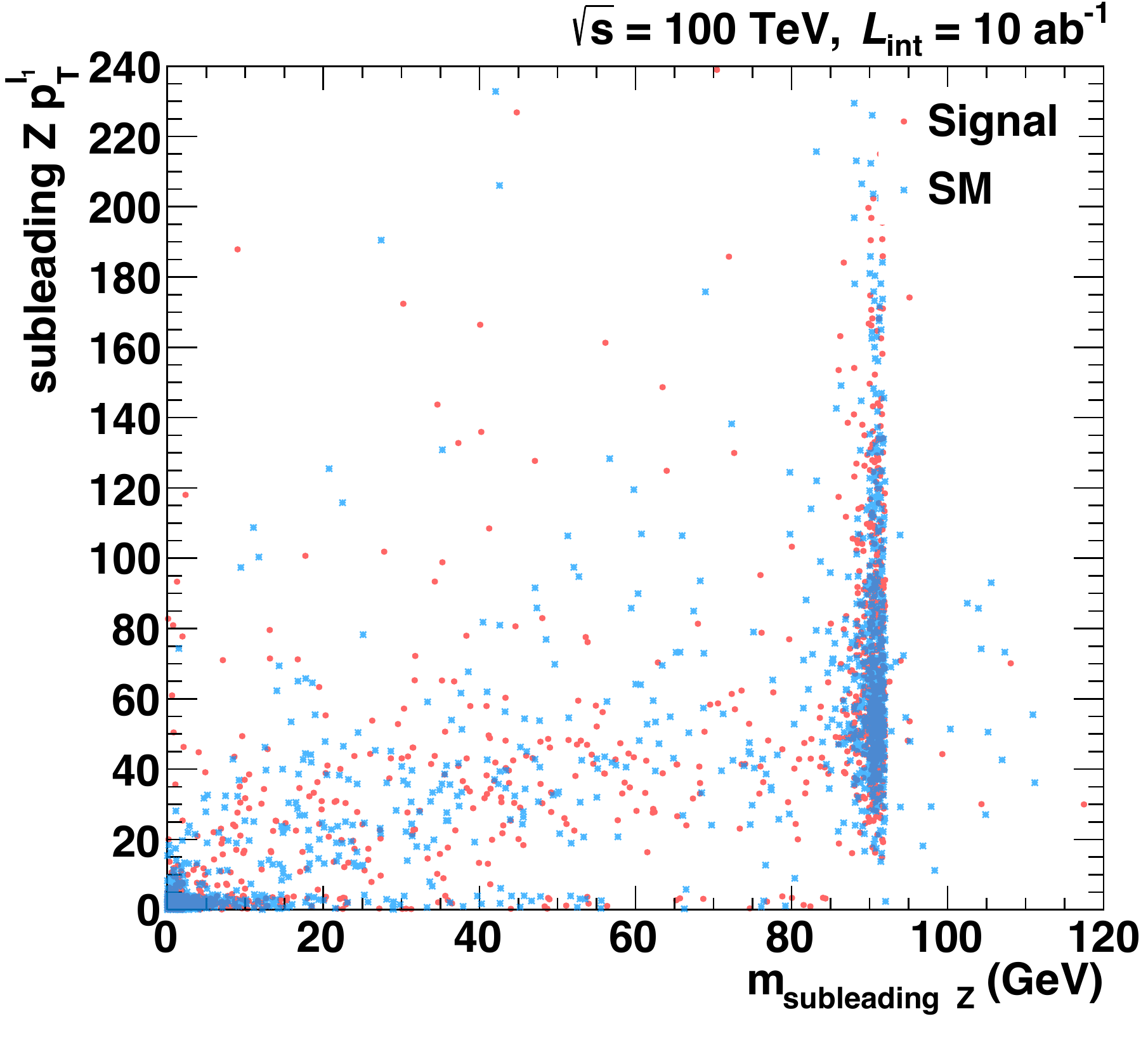}
		\label{fig:e_subLeadingZmassPTl1cut0}
    	}
    	\subfigure[]{
    		\includegraphics[width=0.48\textwidth]{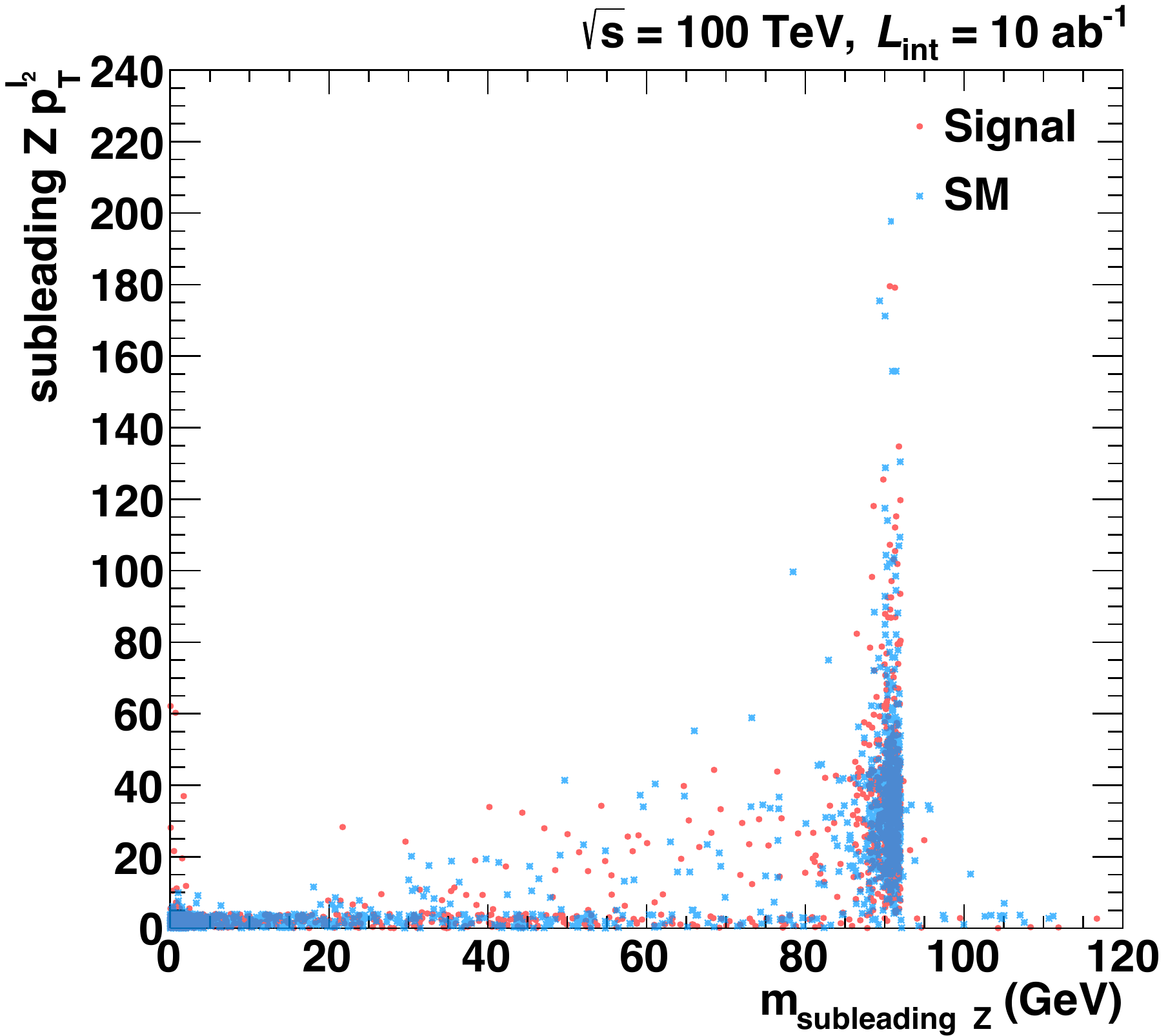}
		\label{fig:e_subLeadingZmassPTl2cut0}
    	}
		
     \caption{Transverse momentum  distributions  of  (a) leading lepton (b) subleading lepton of the $leading \, Z$ boson vs its invariant mass and transverse momentum distributions  of (c) leading lepton (d) subleading lepton of the $subleading \, Z$ boson vs its invariant mass in the  4$e$ channel.}
	\label{fig:e_leadingSubLeadingZmassPTcut0}
\end{figure}

\begin{figure}[!htb] 
 	\centering 
	\subfigure[ ]{
    		\includegraphics[width=0.48\textwidth]{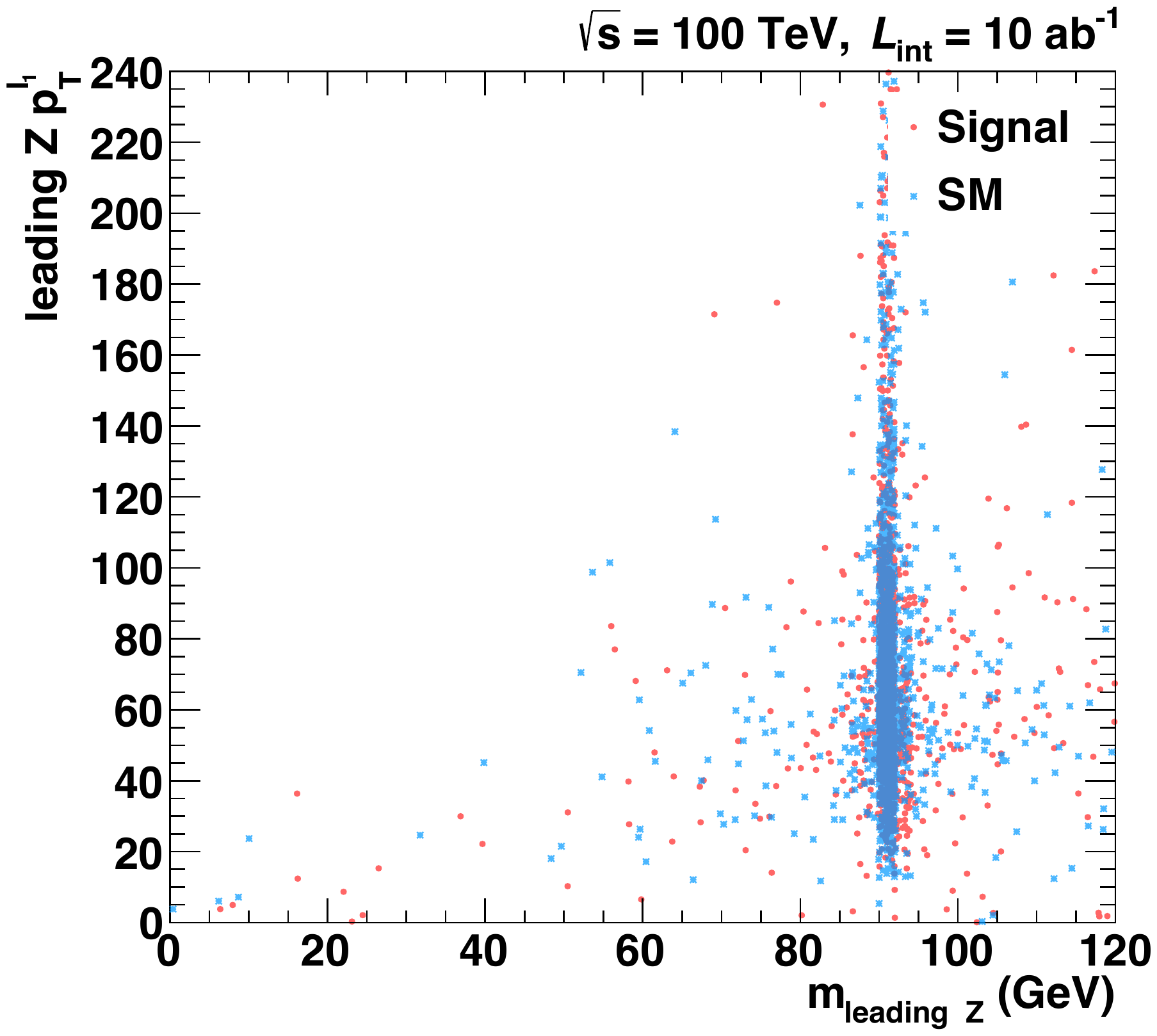}
		\label{fig:mu_leadingZmassPTl1cut0}
    	}
    	\subfigure[]{
    		\includegraphics[width=0.48\textwidth]{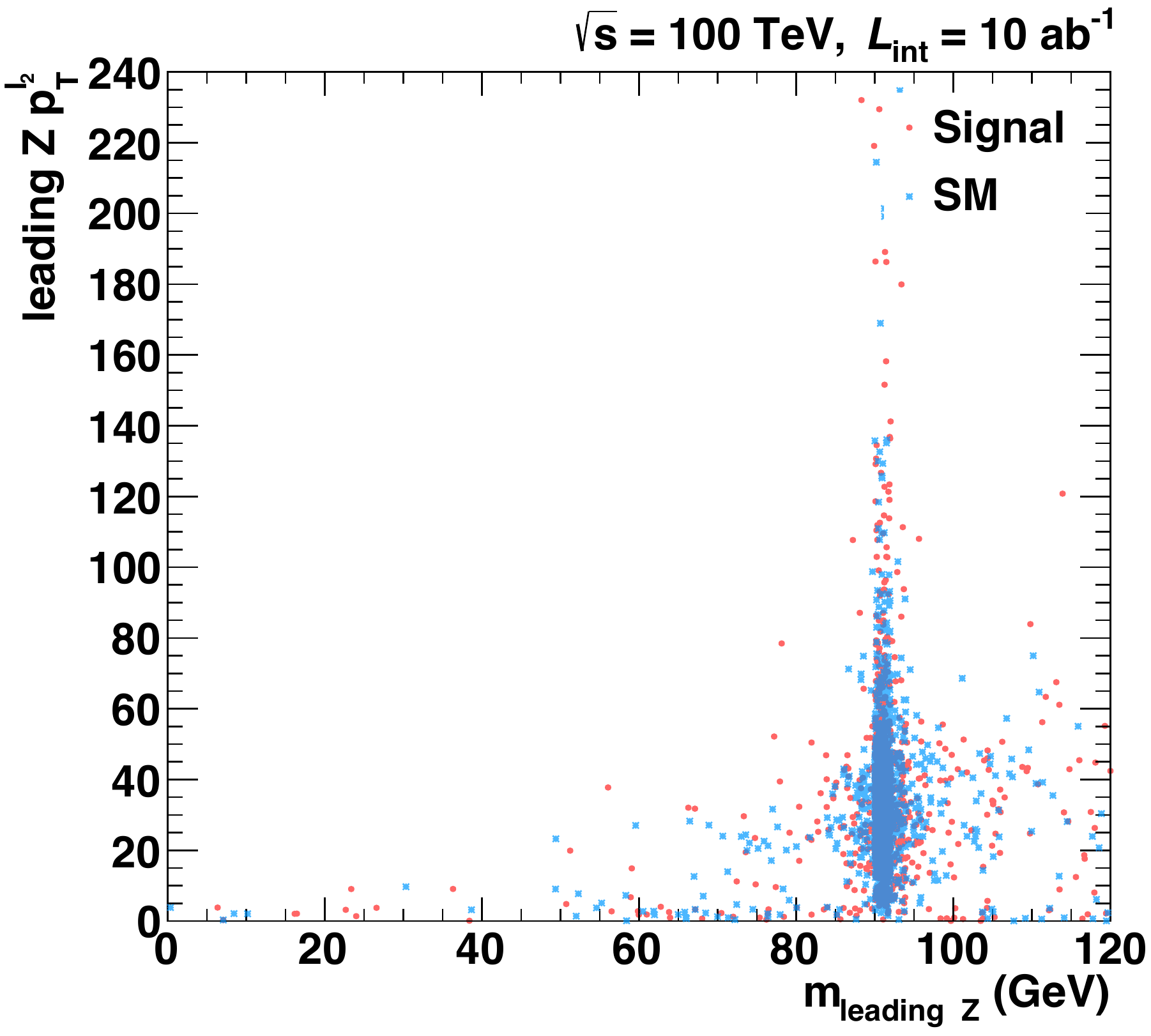}
		\label{fig:mu_leadingZmassPTl2cut0}
    	}
    	\subfigure[]{
    		\includegraphics[width=0.48\textwidth]{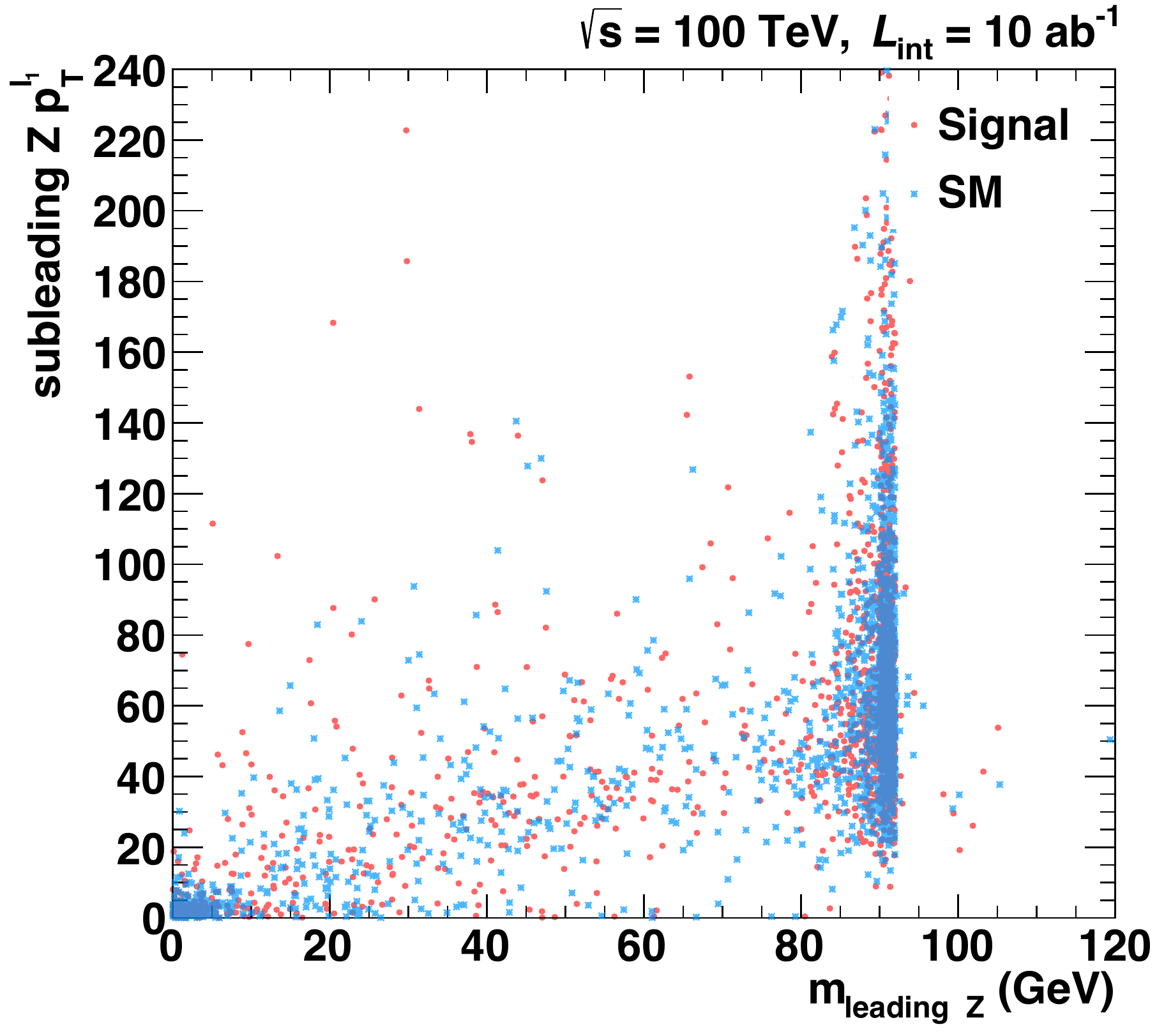}
		\label{fig:mu_subLeadingZmassPTl1cut0}
    	}
    	\subfigure[]{
    		\includegraphics[width=0.48\textwidth]{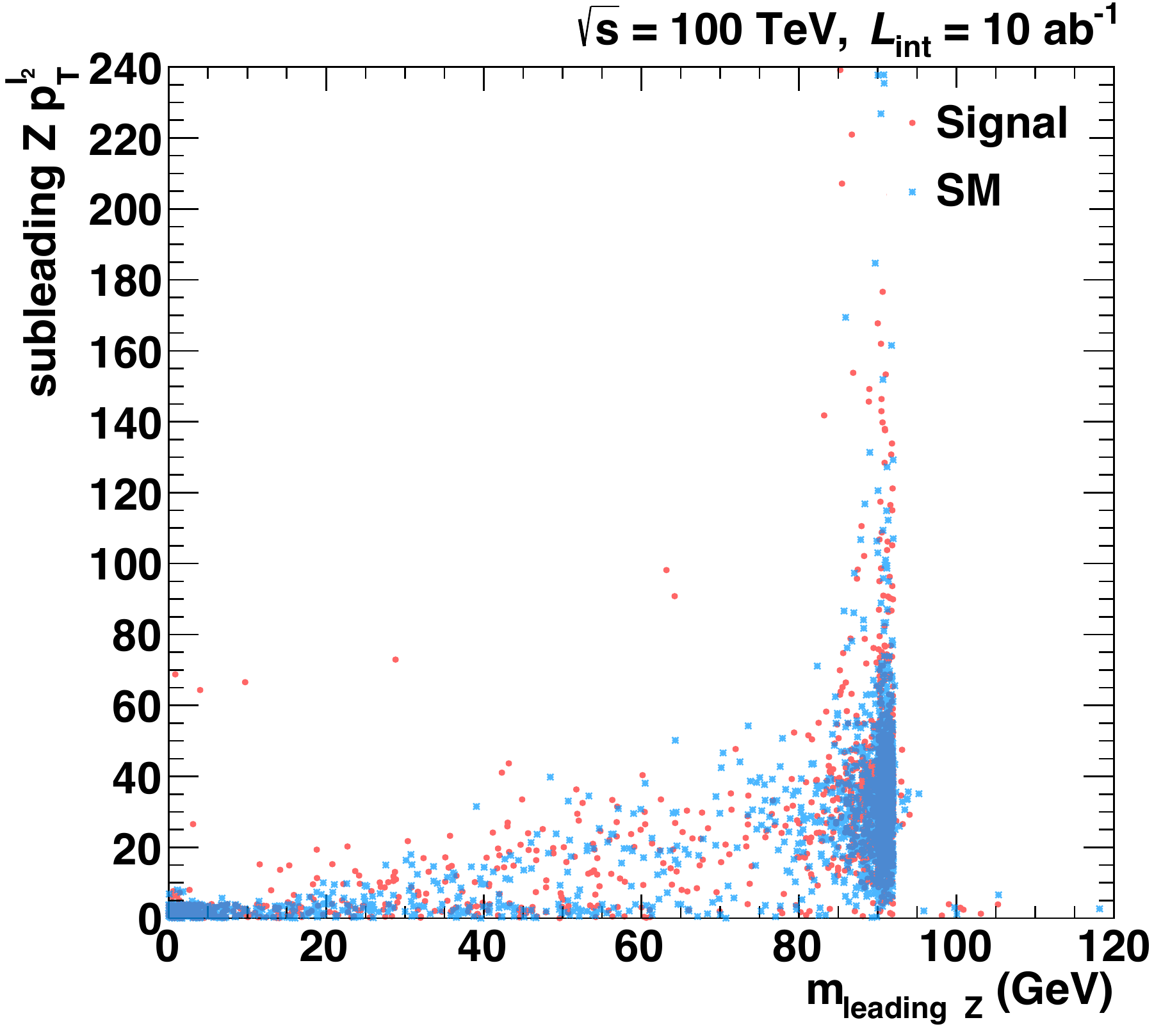}
		\label{fig:mu_subLeadingZmassPTl2cut0}
    	}
		
     \caption{Transverse momentum  distributions  of  (a) leading lepton (b) subleading lepton of the $leading \, Z$ boson vs its invariant mass and transverse momentum distributions  of (c) leading lepton (d) subleading lepton of the $subleading \, Z$ boson vs its invariant mass in the 4$\mu$ channel.}
	\label{fig:mu_leadingSubLeadingZmassPTcut0}
\end{figure}

\begin{figure}[!htb] 
 	\centering 
	\subfigure[]{
    		\includegraphics[width=0.48\textwidth]{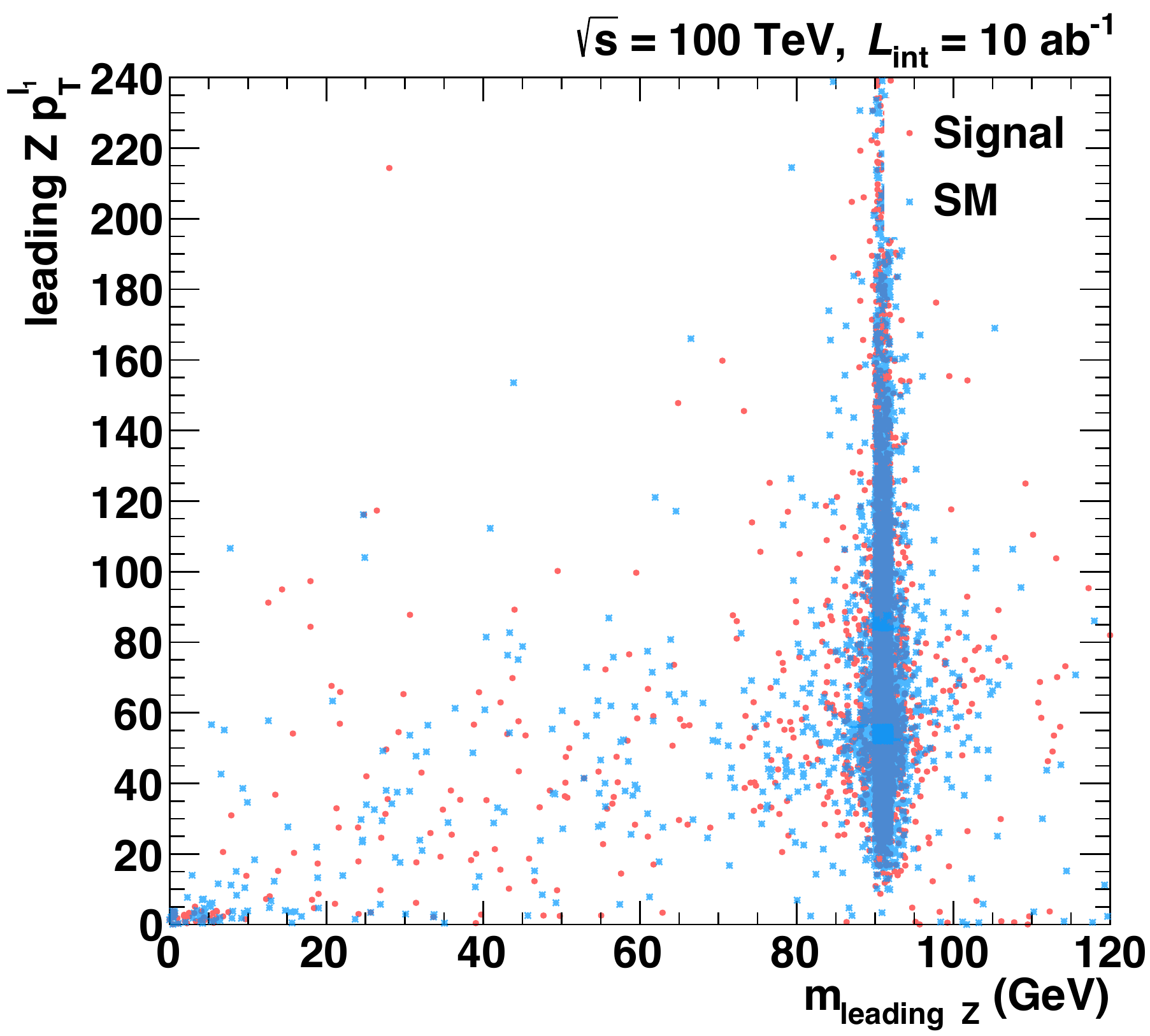}
		\label{fig:eMu_leadingZmassPTl1cut0}
    	}
    	\subfigure[]{
    		\includegraphics[width=0.48\textwidth]{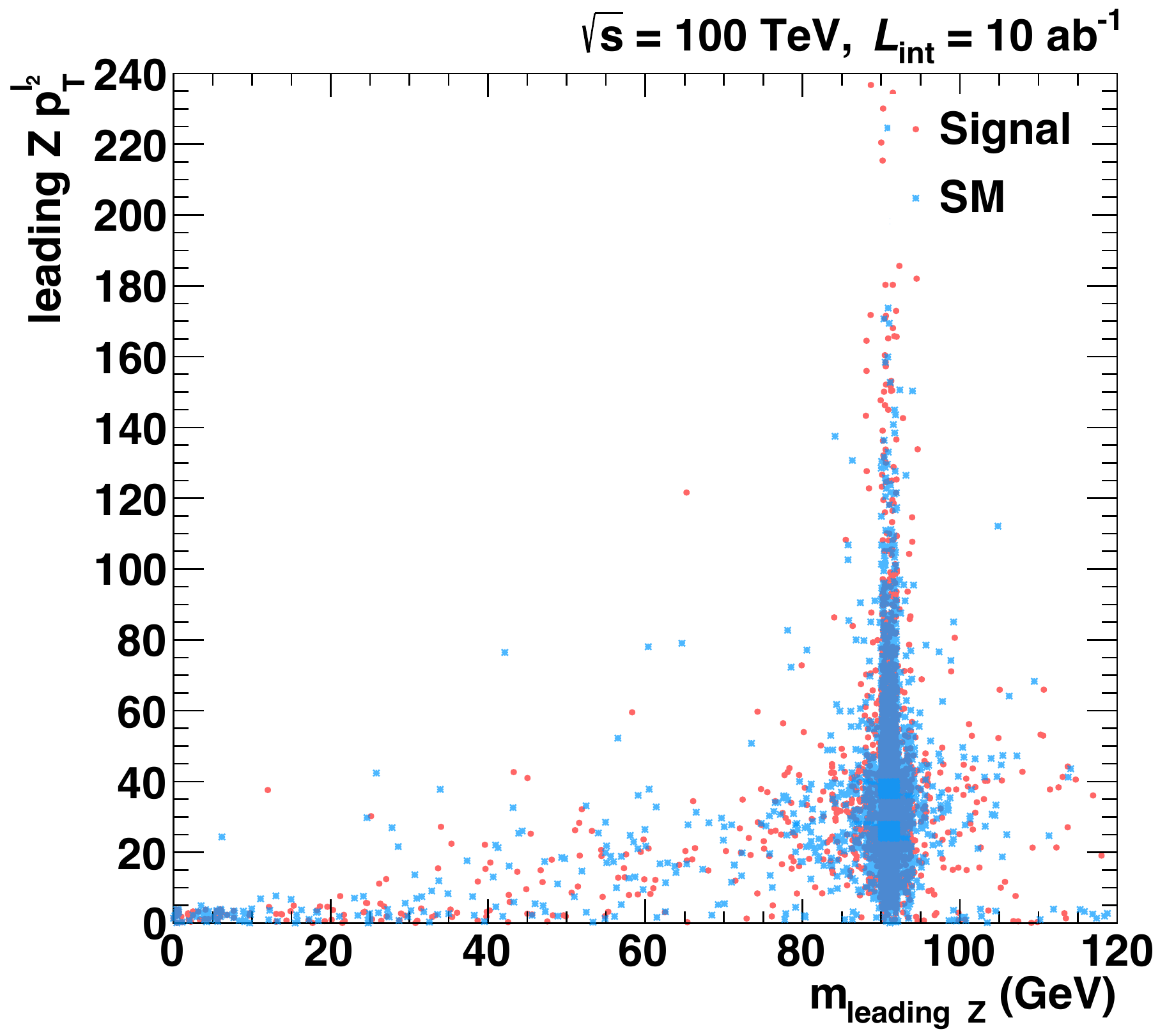}
		\label{fig:eMu_leadingZmassPTl2cut0}
    	}
    	\subfigure[]{
    		\includegraphics[width=0.48\textwidth]{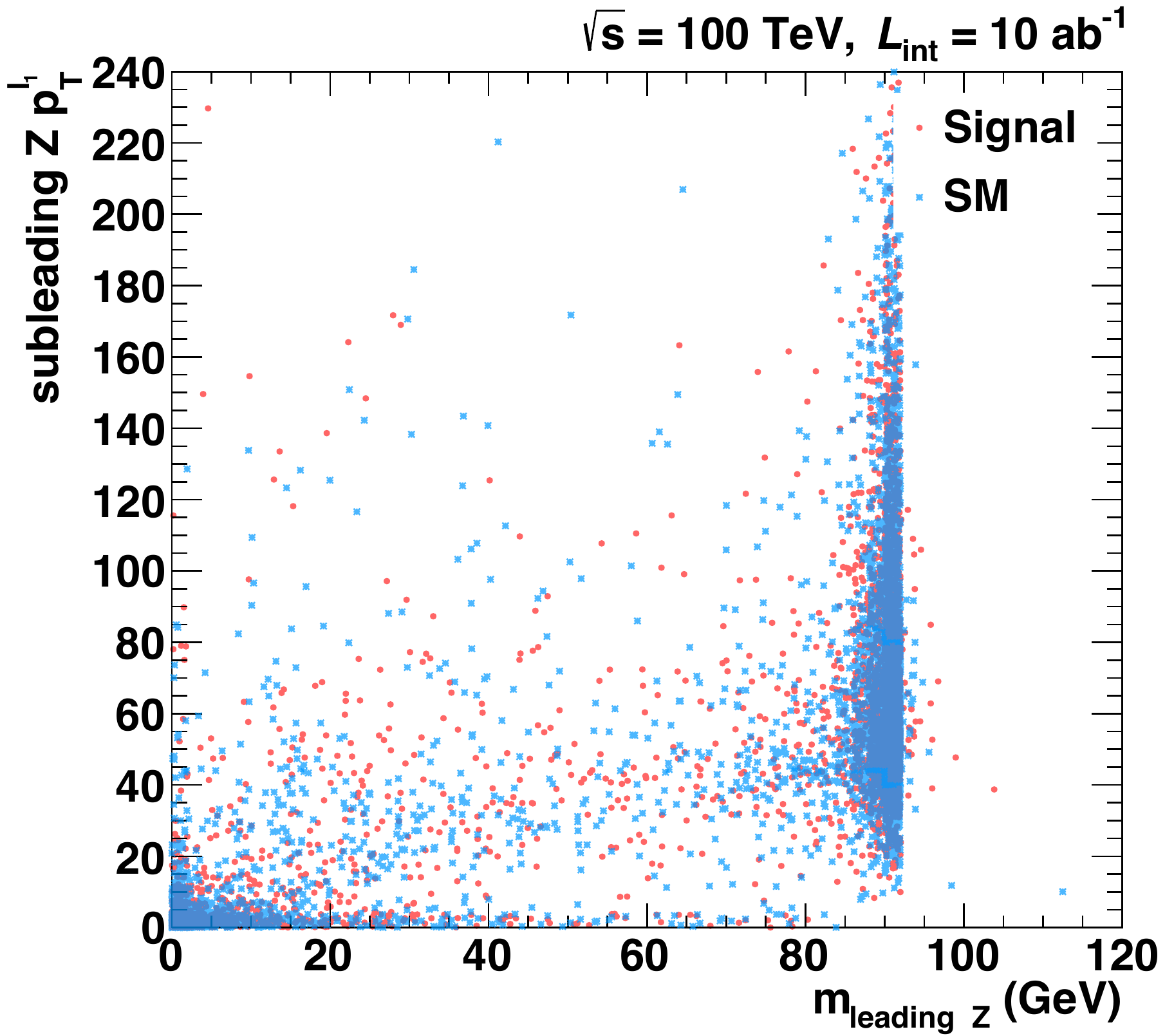}
		\label{fig:eMu_subLeadingZmassPTl1cut0}
    	}
    	\subfigure[]{
    		\includegraphics[width=0.48\textwidth]{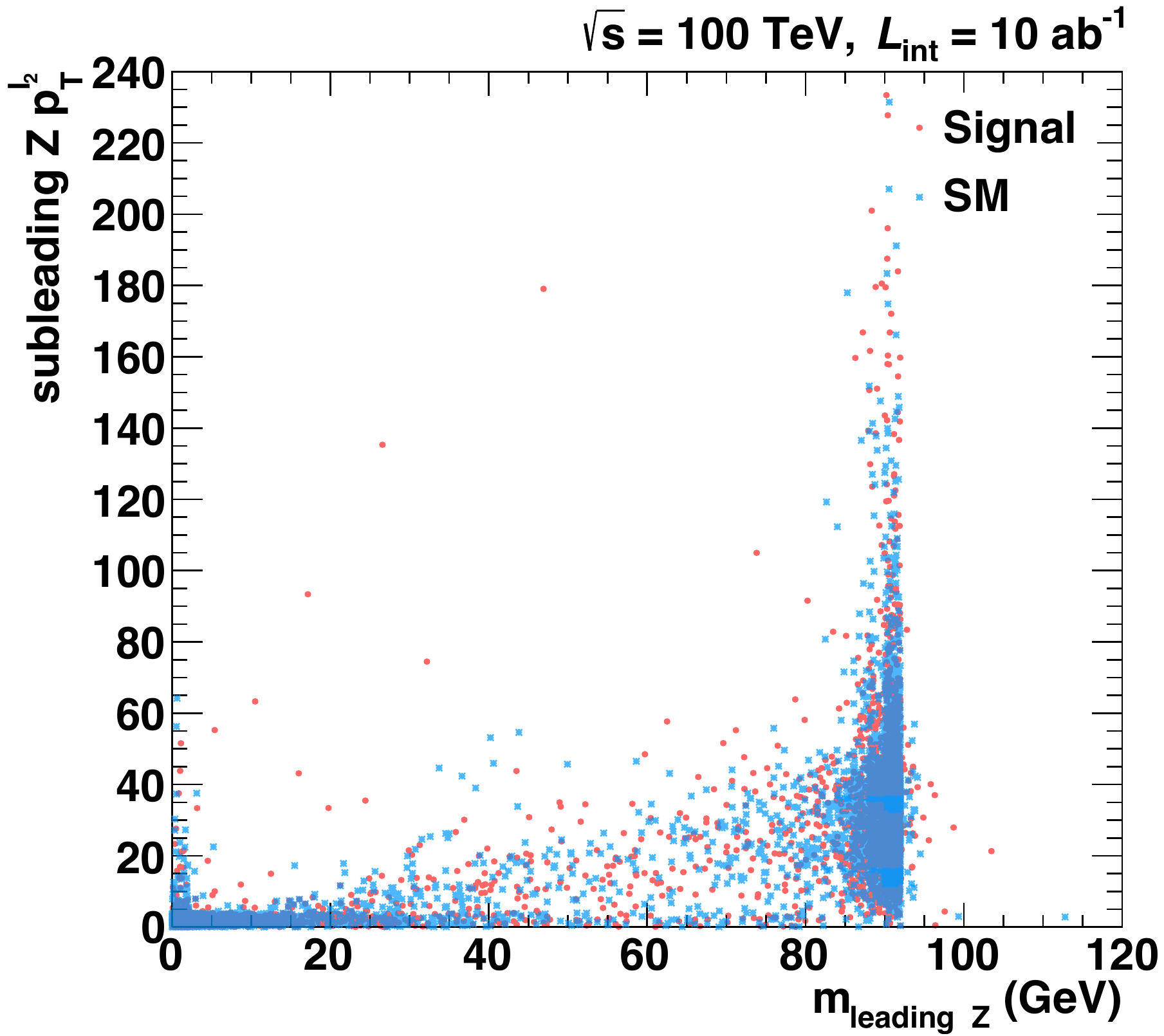}
		\label{fig:eMu_subLeadingZmassPTl2cut0}
    	}
     \caption{Transverse momentum  distributions  of  (a) leading  lepton (b) subleading lepton of the $leading \, Z$ boson vs its invariant mass and transverse momentum distributions  of (c) leading lepton (d) subleading  lepton of the $subleading \, Z$ boson vs its invariant mass in the 2$e$2$\mu$ channel.}
	\label{fig:eMu_leadingSubLeadingZmassPTcut0}
\end{figure}

In order to see the region where the signal can be enhanced we plotted the transverse momentum of leptons ($p_{T}^{\ell^{1}}$, $p_{T}^{\ell^2}$) versus the reconstructed invariant mass of the $leading$ and $subleading\, Z$ as shown in Fig.~\ref{fig:e_leadingSubLeadingZmassPTcut0}, Fig.~\ref{fig:mu_leadingSubLeadingZmassPTcut0} and Fig.~\ref{fig:eMu_leadingSubLeadingZmassPTcut0} for 4$e$, 4$\mu$ and 2$e$2$\mu$ channels, respectively. $p_{T}^{\ell^{1}}$ is labelled as the highest-$p_{T}$ lepton in both $leading$ and $subleading \,Z$. 
The cut for  highest $p_{T}^{\ell^{1}}$ lepton is greater than 20 GeV, and for the subleading lepton is $p_{T}^{\ell^2}>$ 12 GeV  (10 GeV)  in the $leading \, Z$, while the remaining leptons in the $subleading \, Z$ must have $p_{T}^{\ell^1, \ell^2}>$ 5 GeV for electrons (muons). 

\begin{figure}[!htb] 
 	\centering 
	\subfigure[]{
    		\includegraphics[width=0.48\textwidth]{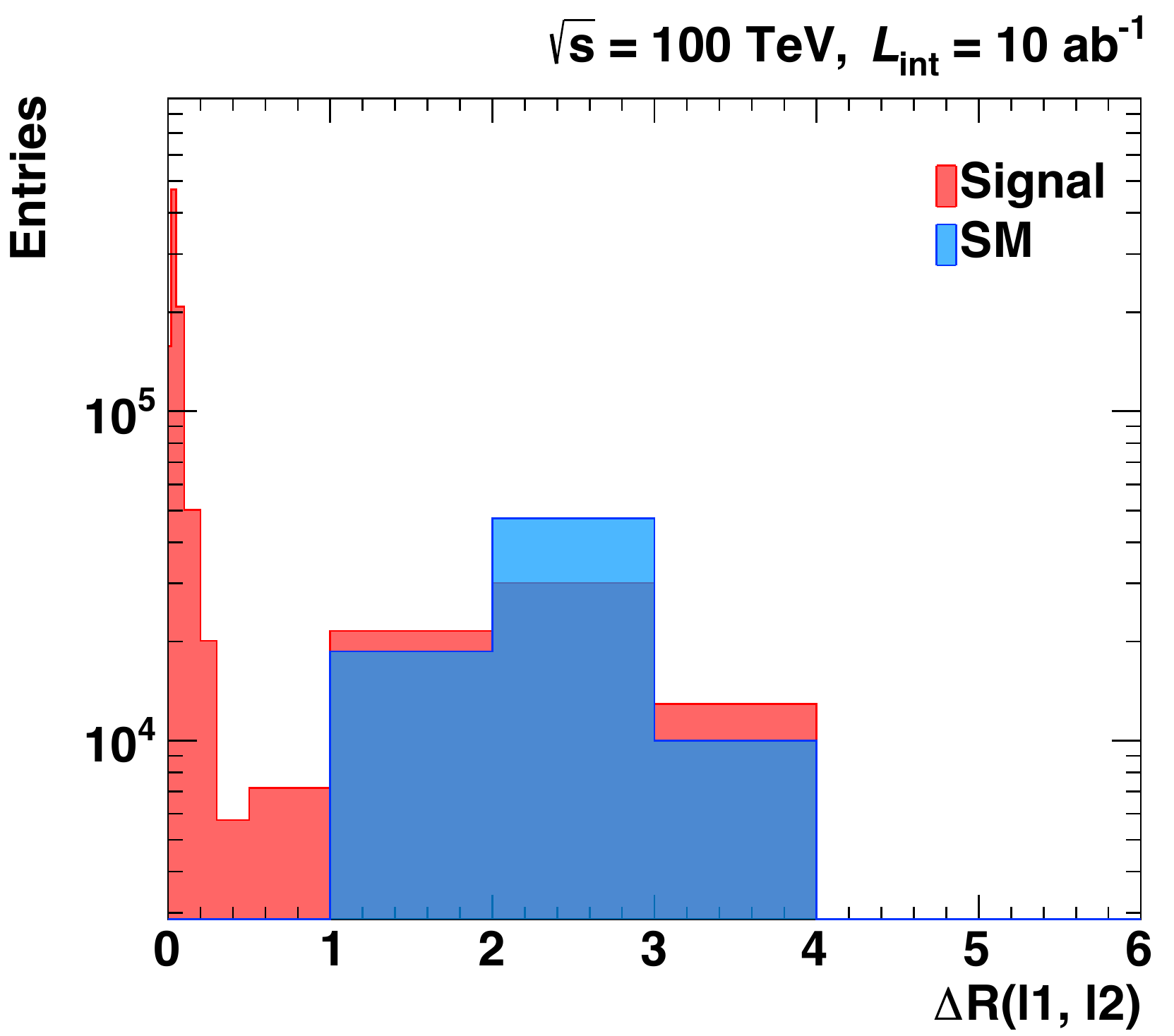}
		\label{fig:e_leadingZdRl1l2}
    	}
    	\subfigure[]{
    		\includegraphics[width=0.48\textwidth]{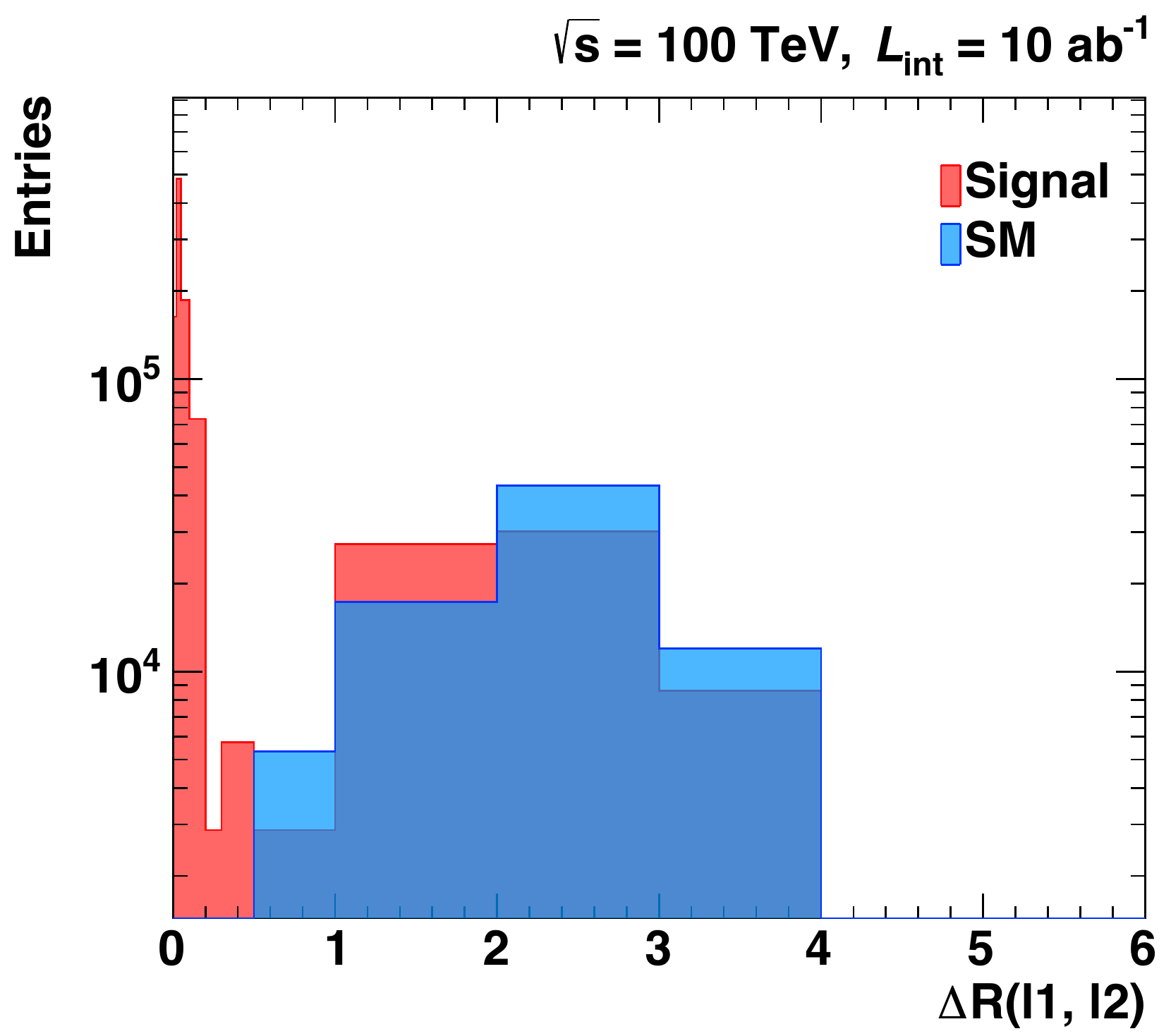}
		\label{fig:e_subleadingZdRl1l2}
    	}
    	\subfigure[ ]{
    		\includegraphics[width=0.48\textwidth]{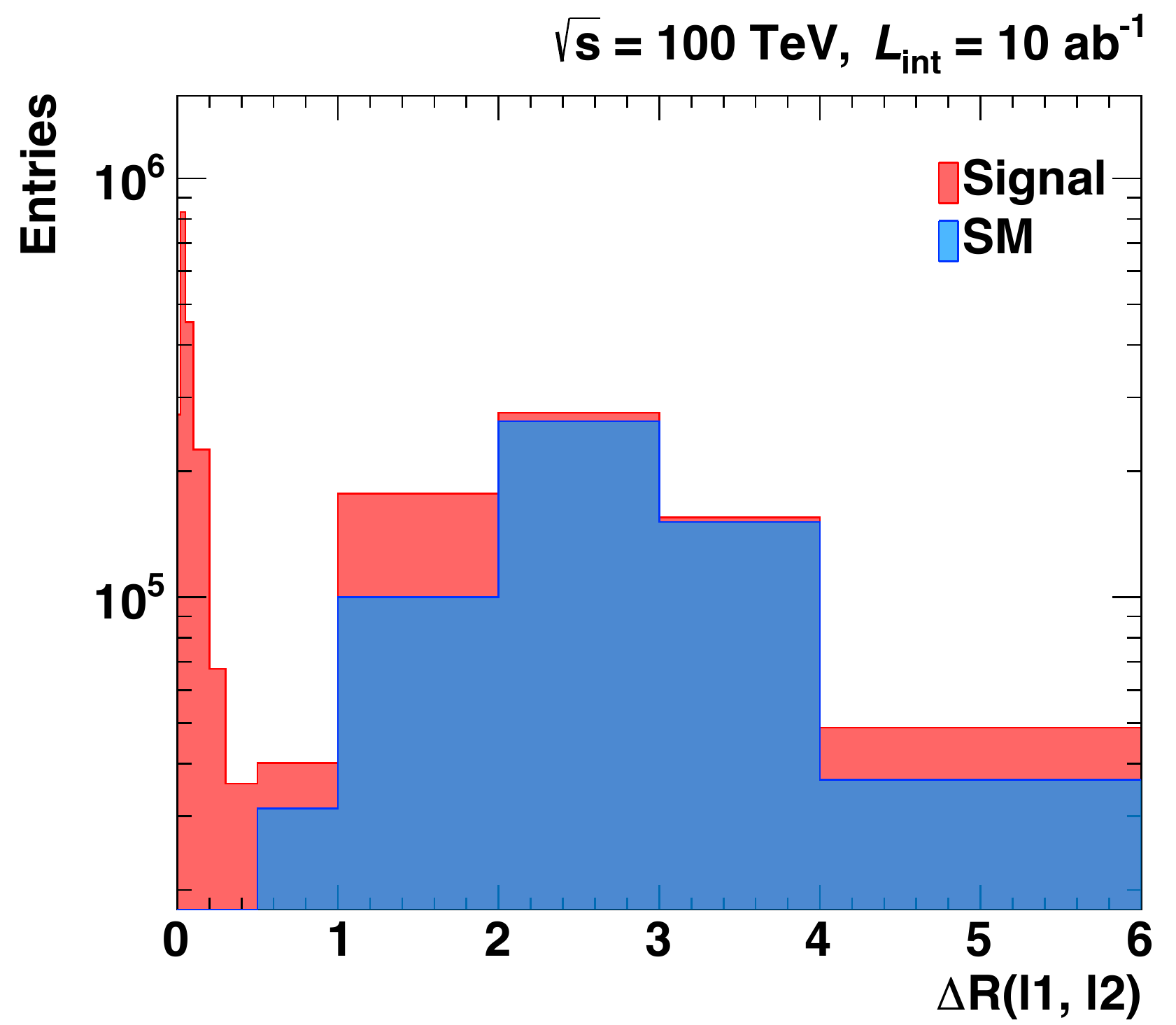}
		\label{fig:mu_leadingZdRl1l2}
    	}
    	\subfigure[]{
    		\includegraphics[width=0.48\textwidth]{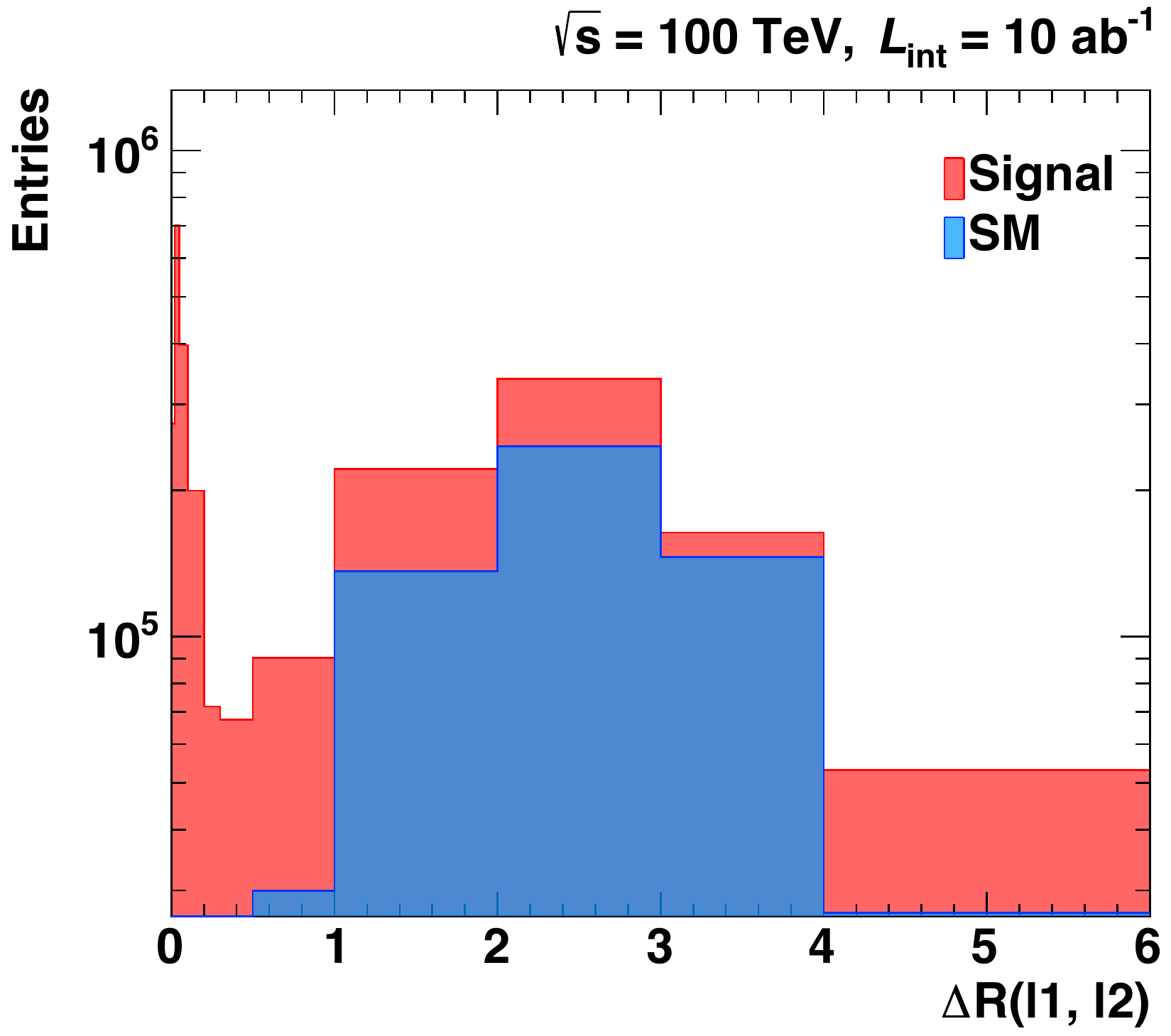}
		\label{fig:mu_subleadingZdRl1l2}
    	}
	\subfigure[]{
    		\includegraphics[width=0.48\textwidth]{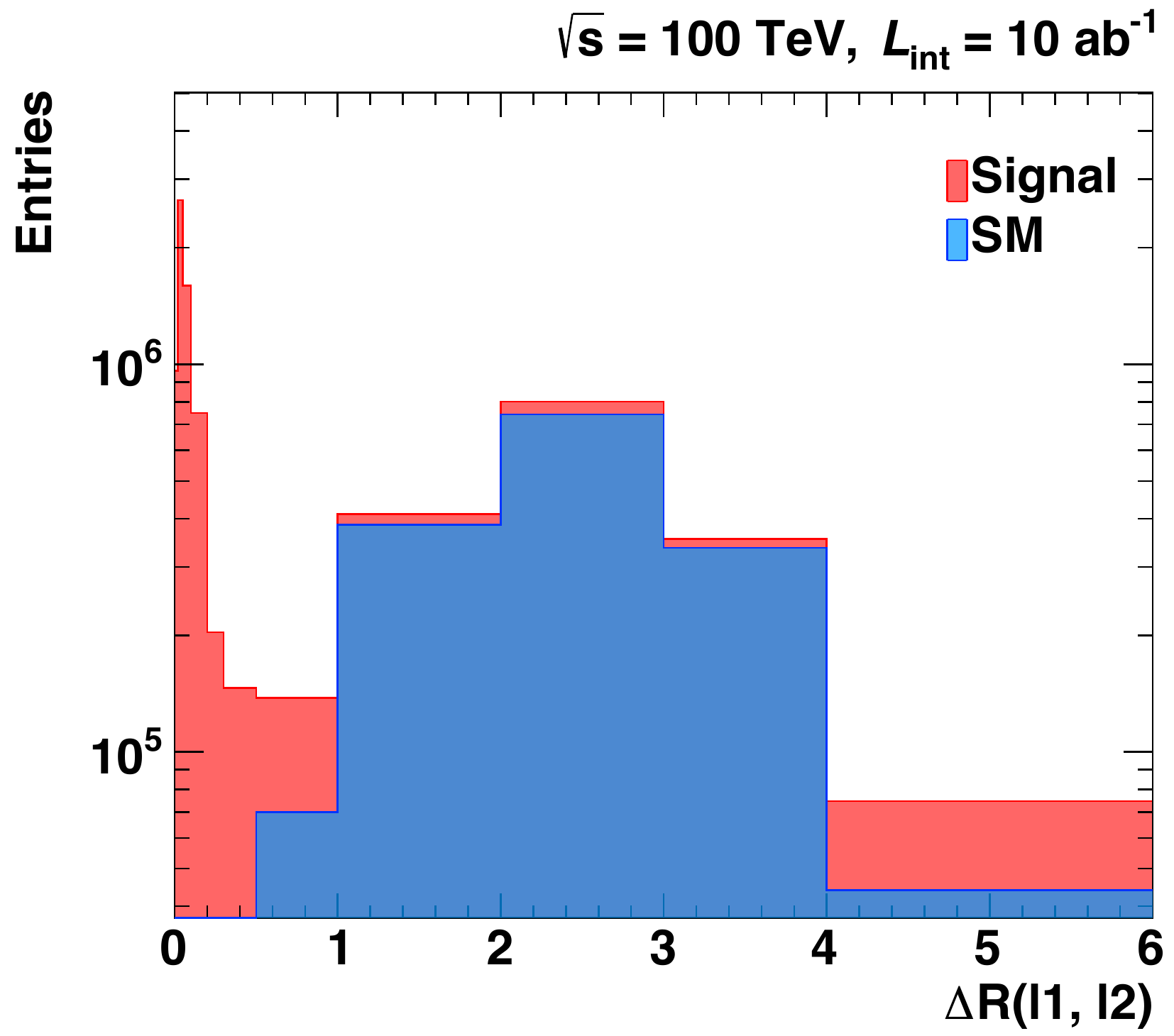}
		\label{fig:eMu_leadingZdRl1l2}
    	}
    	\subfigure[]{
    		\includegraphics[width=0.48\textwidth]{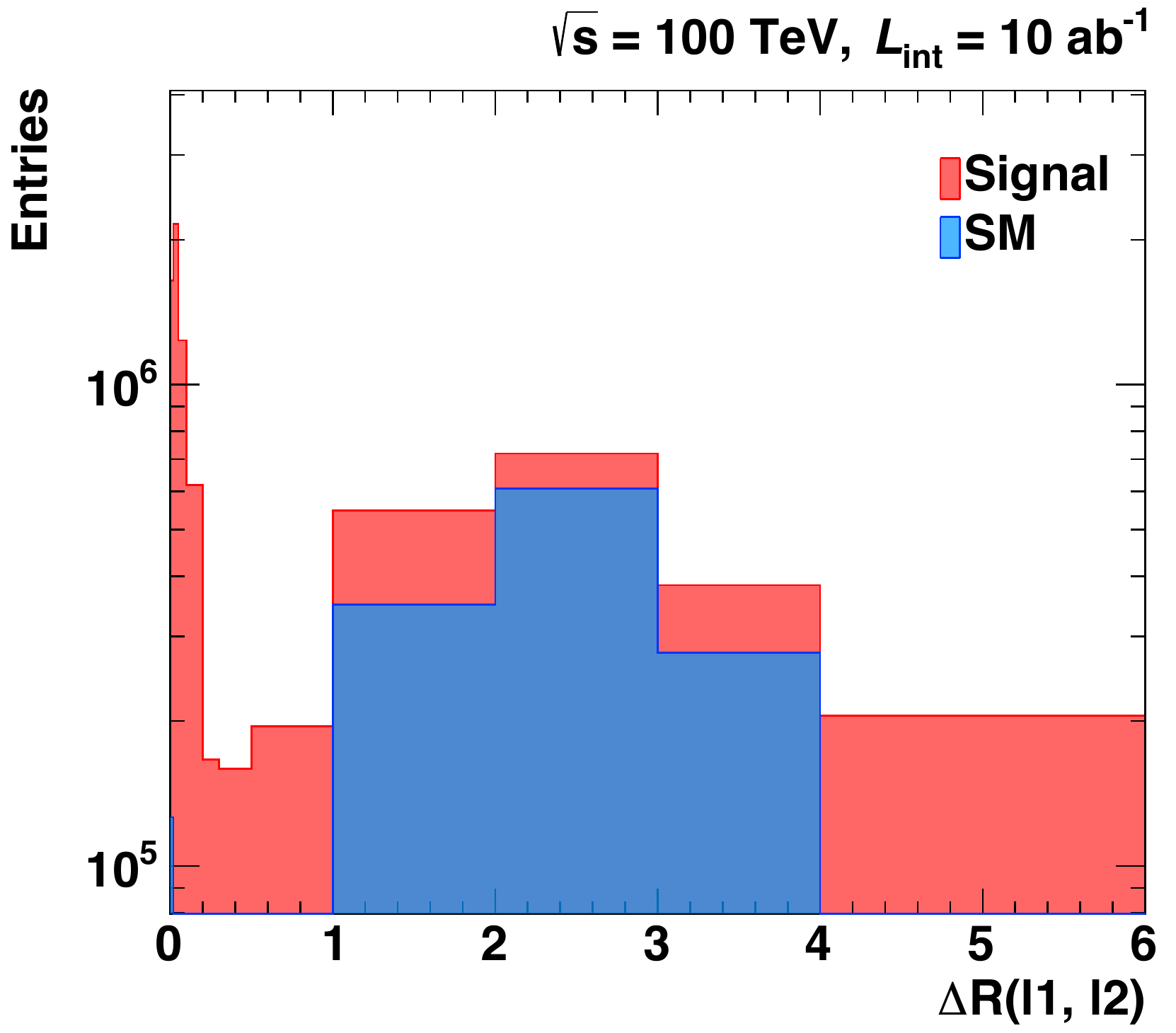}
		\label{fig:eMu_subleadingZdRl1l2}
    	}
		
     \caption{$\Delta R$ distributions between leptons of candidate Z boson pairs in the 4$e$ (a, b), 4$\mu$ (c, d) and 2$e$2$\mu$ (e, f) channels. Fig. (a), (c) and (e) for $leading\, Z$  and Fig. (b), (d) and (f) for $subleading\, Z$ bosons.}
	\label{fig:deltaR}
\end{figure}

The pseudo-rapidity cuts of all leptons are applied as $|\eta^{\ell} |<$ 2.5. The distance $\Delta R (\ell^{1} , \ell^{2} )$ between leptons in $\eta$-$\phi$ plane is evaluated by the function 
\begin{equation}
\Delta R (\ell^{1} , \ell^{2} ) = \sqrt{(\eta^{\ell^{1}} - \eta^{\ell^{2}}  )^{2} + (\phi^{\ell^{1}} - \phi^{\ell^{2}})^{2}}
\label{eqn:dR}
\end{equation}
and plotted in Fig.~\ref{fig:deltaR}.
This figure shows $\Delta R$ distributions between two leptons of leading and subleading Z in the first and second column. Each row corresponds to different decay channel aligned for 4$e$, 4$\mu$ and 2$e$2$\mu$, respectively. 
In order to meet the detector requirement, we applied a cut for all leptons are separated from each others by imposing $\Delta R (\ell^{1} , \ell^{2} ) >$ 0.02.

The $pp \rightarrow ZZ$ sensitivity is estimated by using events  where
a further cut is applied for both invariant mass of $leading \, Z$ and $subleading \, Z$ bosons must be within the range 80 $< m_{leading \,  Z} <$ 100 GeV and 60 $< m_{subleading \, Z} <$ 110 GeV, respectively. This ranges were chosen to keep most of the decays in the resonance while removing mostly other processes with 4$\ell$ final states.
 Decays of the $Z$ bosons to $\tau$ leptons with subsequent decays to electrons and muons are heavily suppressed by requirements on lepton $p_{T}$.
The cut flow steps in the analysis for selecting the events are summarized in Table~\ref{tab:tab2}.

\begin{table}[!htp]
\caption{Preselection and a set of cuts for the analysis of signal and background
events. \label{tab:tab2}}

\centering%
\resizebox{\textwidth}{!}{
\begin{tabular}{ll}
\hline 
Cuts  &   Definition \tabularnewline
\hline 
Cut-0  &   Preselection: $N_{\ell_{(e, \mu)}}>=$ 4 and \tabularnewline
           &     two same-flavor opposite-charge lepton pairs \tabularnewline
Cut-1  &   Dileptons minimizing  $| m^{a}_{\ell\ell} - m_{Z} | + | m^{b}_{\ell\ell} - m_{Z} | $ are taken as \tabularnewline
           &   Z boson pair candidates \tabularnewline
Cut-2  &   Transverse momentum: $p_{T}^{\ell^{1}} >$ 20 GeV, $p_{T}^{\ell^{2}} >$ 12 GeV (10 GeV) \tabularnewline
	  &   for $e \, (\mu)$ and $p_{T}^{\ell^{3,4}} >$ 5 GeV \tabularnewline

Cut-3  &  Pseudo-rapidity: $|\eta^{\ell}|<$ 2.5 \tabularnewline
Cut-4  &  $\Delta R >$ 0.02 between all leptons \tabularnewline
Cut-5  &  Invariant mass: 80 $<M_{inv}^{rec}$(leading Z)$<$ 100 GeV and  \tabularnewline
          & 60 $<M_{inv}^{rec}$(subleading Z)$<$ 110 GeV \tabularnewline
\hline 
\end{tabular}
}
\end{table}

\begin{figure}[!htb] 
 	\centering 
	\subfigure[]{
 		  \includegraphics[width=0.465\textwidth]{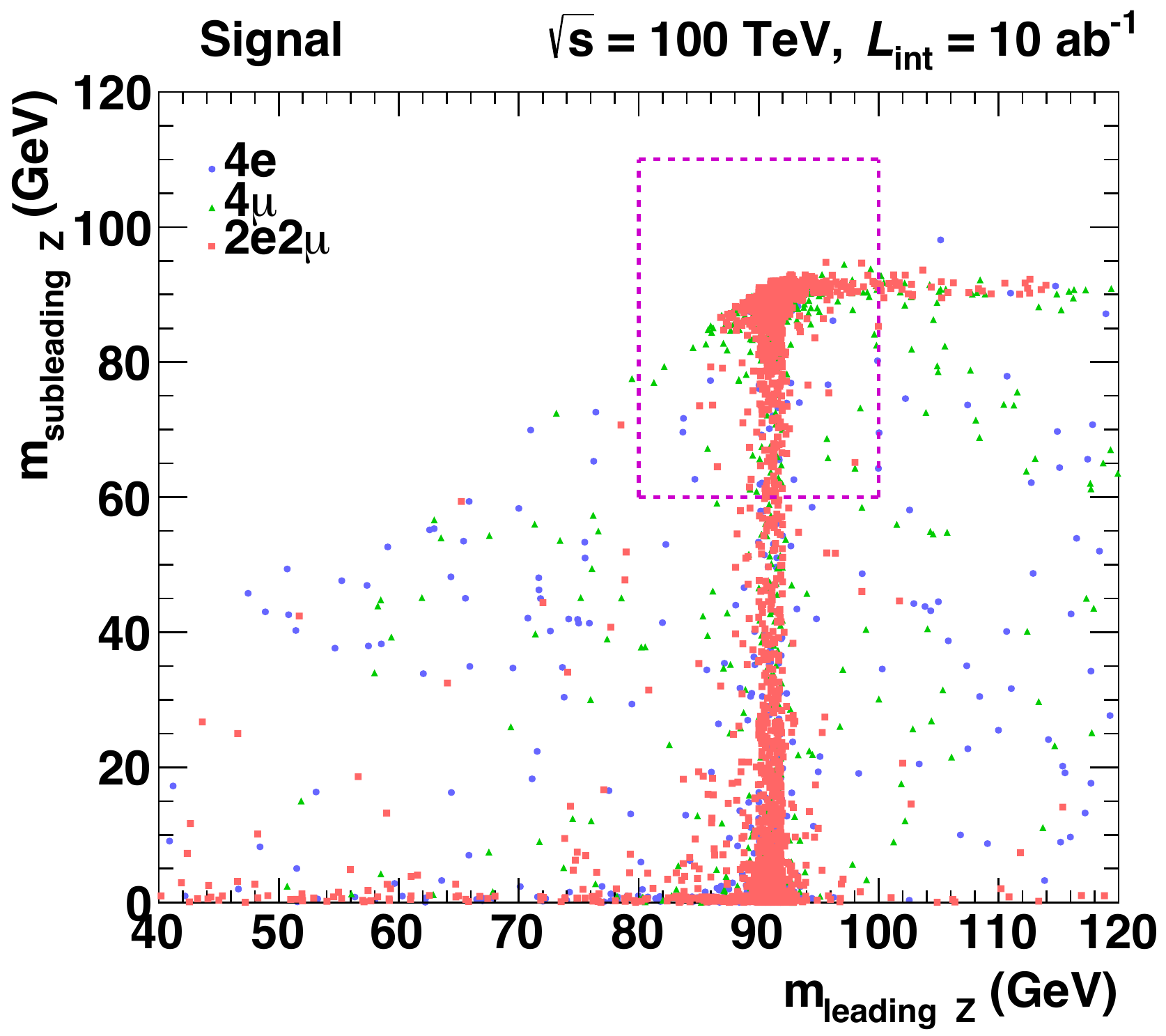}
		  }
		  \subfigure[]{
		    \includegraphics[width=0.465\textwidth]{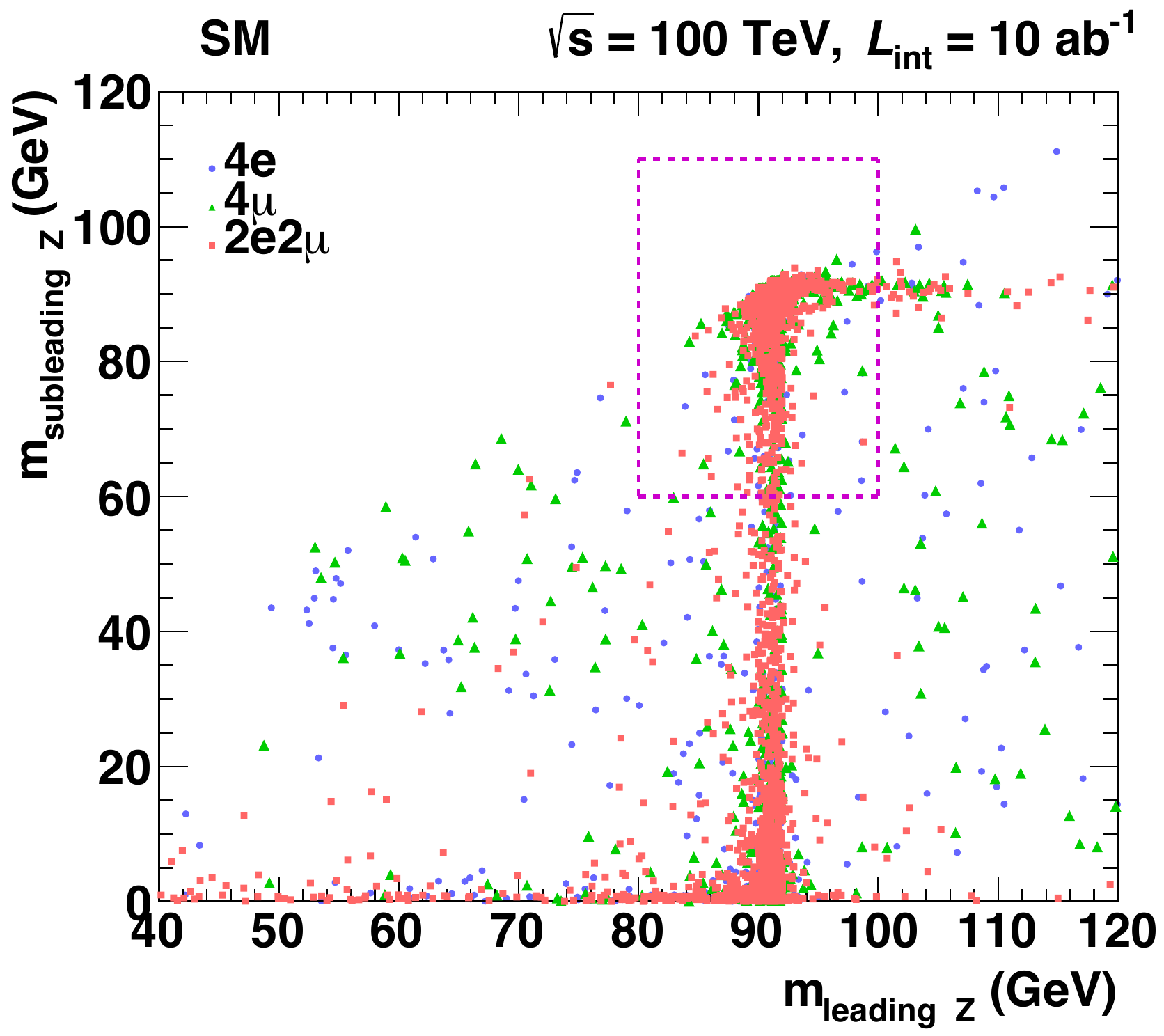}

		  }
     \caption{Invariant mass distributions  of $subleading \, Z$ boson versus $leading \, Z$ boson  for signal (left) and main background (right)  with an individual markers for each 4$e$, 4$\mu$ and 2$e$2$\mu$ channels.}
	\label{fig:all_channel_LeadingSubLeadingZmassCut0}
\end{figure}

After applying the kinematical cuts discussed above, the reconstructed invariant mass of the $leading \, Z$ boson candidates, and a scatter plot showing the correlation between $subleading \, Z$ boson versus $leading \, Z$ boson in simulated events, are shown in Fig.~\ref{fig:all_channel_LeadingSubLeadingZmassCut0}. 

\begin{table}[!htp]
\caption{The number of events yielded for main background and signal (where  all couplings equal to zero, except $C_{\widetilde{B}W}=5$) of four-lepton events in the mass region 80 $< m_{4\ell} <$ 100 GeV shown for each final state and combined  at FCC-hh with $\mathcal{L}_{int}=$ 10 ab$^{-1}$. \label{tab:eventNumbers} }
\centering{}%

\begin{tabular}{lcccl } 
\hline 
Channel & Signal & Background   & Total \tabularnewline
\hline 
$4e$          & 16308 & 13991  & 30299  \tabularnewline
$4\mu$      & 32477 & 26850 & 59327  \tabularnewline
$2e2\mu$  & 76404 & 71755 & 148159  \tabularnewline
\hline 
\end{tabular}
\end{table}

\section{Results}\label{sec:results}

To obtain 95\% C.L. limits on the couplings, we apply $\chi^2$ criterion without and with a systematic error. The $\chi^2$ function is defined as follows

\begin{equation}
\label{eqn:chi2def}
\chi^{2} =\sum_i^{n_{bins}}\left(\frac{N_{i}^{NP}-N_{i}^{B}}{N_{i}^{B}\Delta_i}\right)^{2}
\end{equation}
where $N_{i}^{NP}$ is the total number of events in the existence of effective couplings, $N_{i}^{B}$ is total number of events of the corresponding SM backgrounds in $i$th bin of the invariant mass of the quartet-leptons distribution, $\Delta_i=\sqrt{\delta_{sys}^2+ 1 / N_i^B}$ is the combined systematic ($\delta_{sys}$) and statistical errors in each bin. 

\begin{figure}[!htb] 
  \centering 
   	\includegraphics[width=0.8\textwidth]{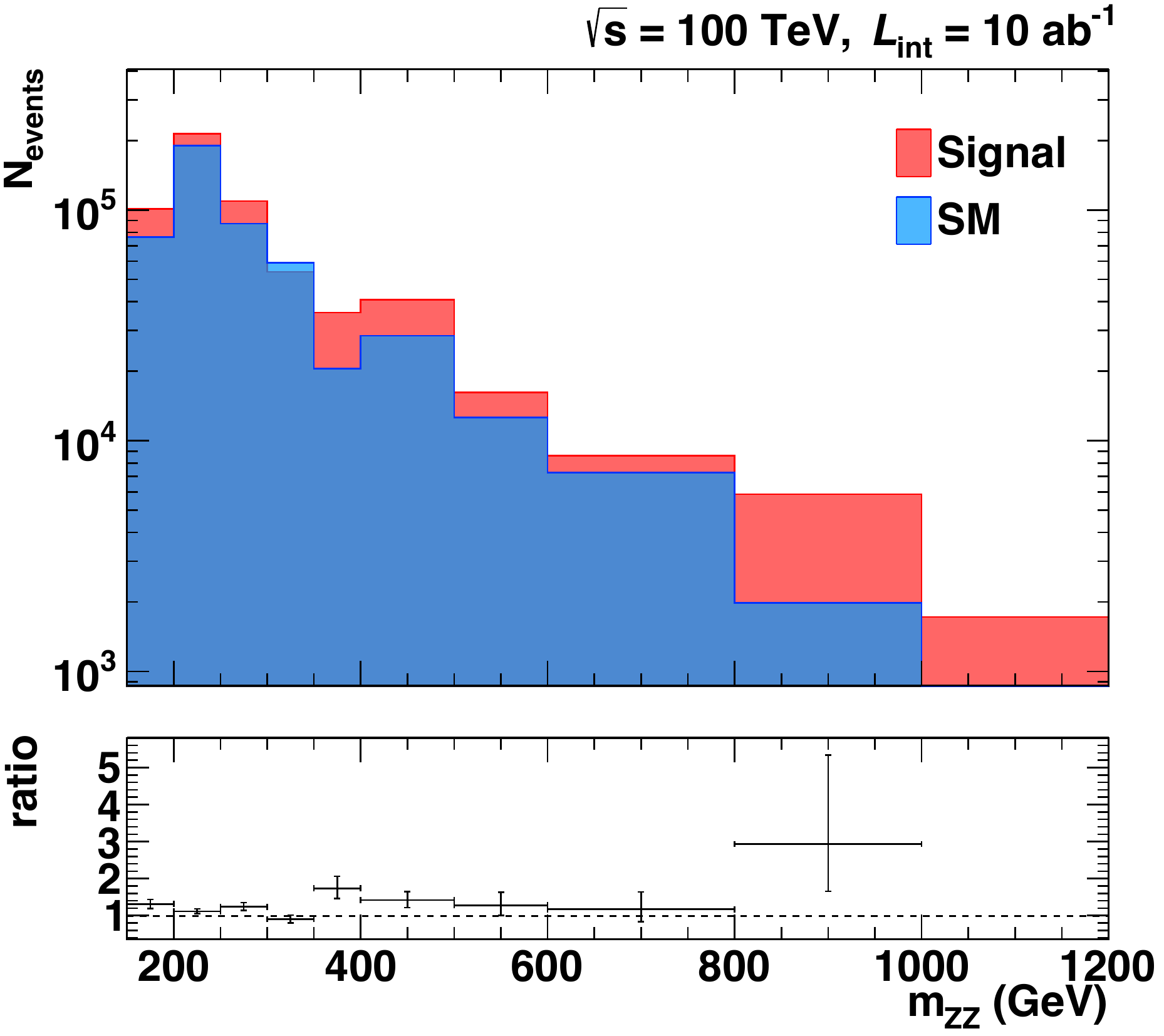}
  \caption{Distributions of the reconstructed four-lepton invariant mass $m_{ZZ}$. In the $m_{ZZ}$ distribution, bin contents are normalized to the bin widths. The lower plot shows the ratio of signal and background int eh bins.} 
   \label{fig:mZZ4l}
\end{figure}

The existence of aTGCs will lead to enhance the yield of events at quadruplet-lepton masses. The distribution of the quadruplet-lepton reconstructed mass of events with both leading and subleading $Z$ bosons in the mass range 60–120 GeV for the unified 4$e$, 4$\mu$, and 2$e$2$\mu$ channels are depicted in Fig.~\ref{fig:mZZ4l}. The limits on probable contributions from aNTGCs are extracted by using this distributions.

\begin{figure}[!htb] 
 	\centering 
		\subfigure[]{
 		  \includegraphics[width=0.465\textwidth]{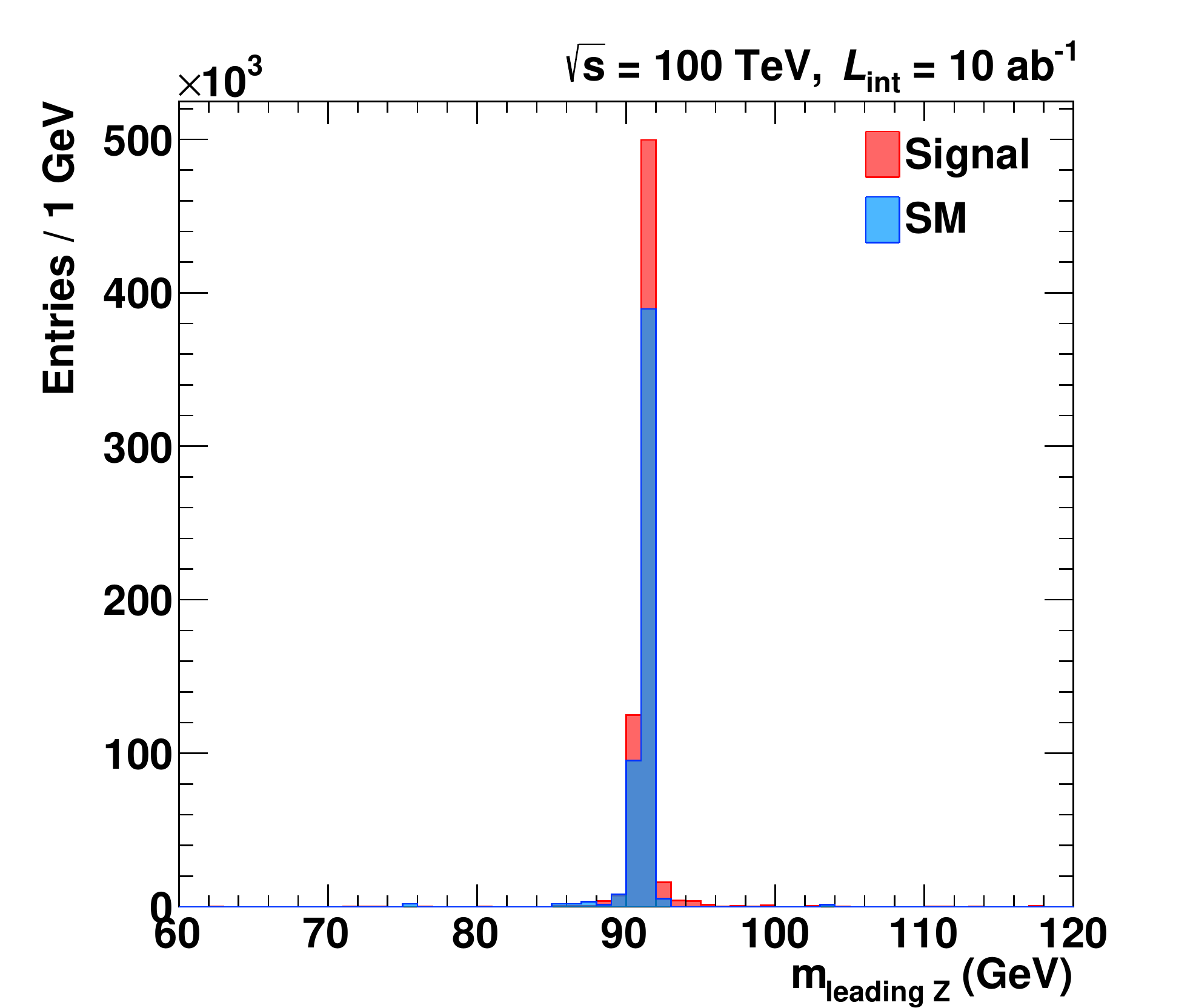}
		  }
		  	\subfigure[]{
			 		  \includegraphics[width=0.465\textwidth]{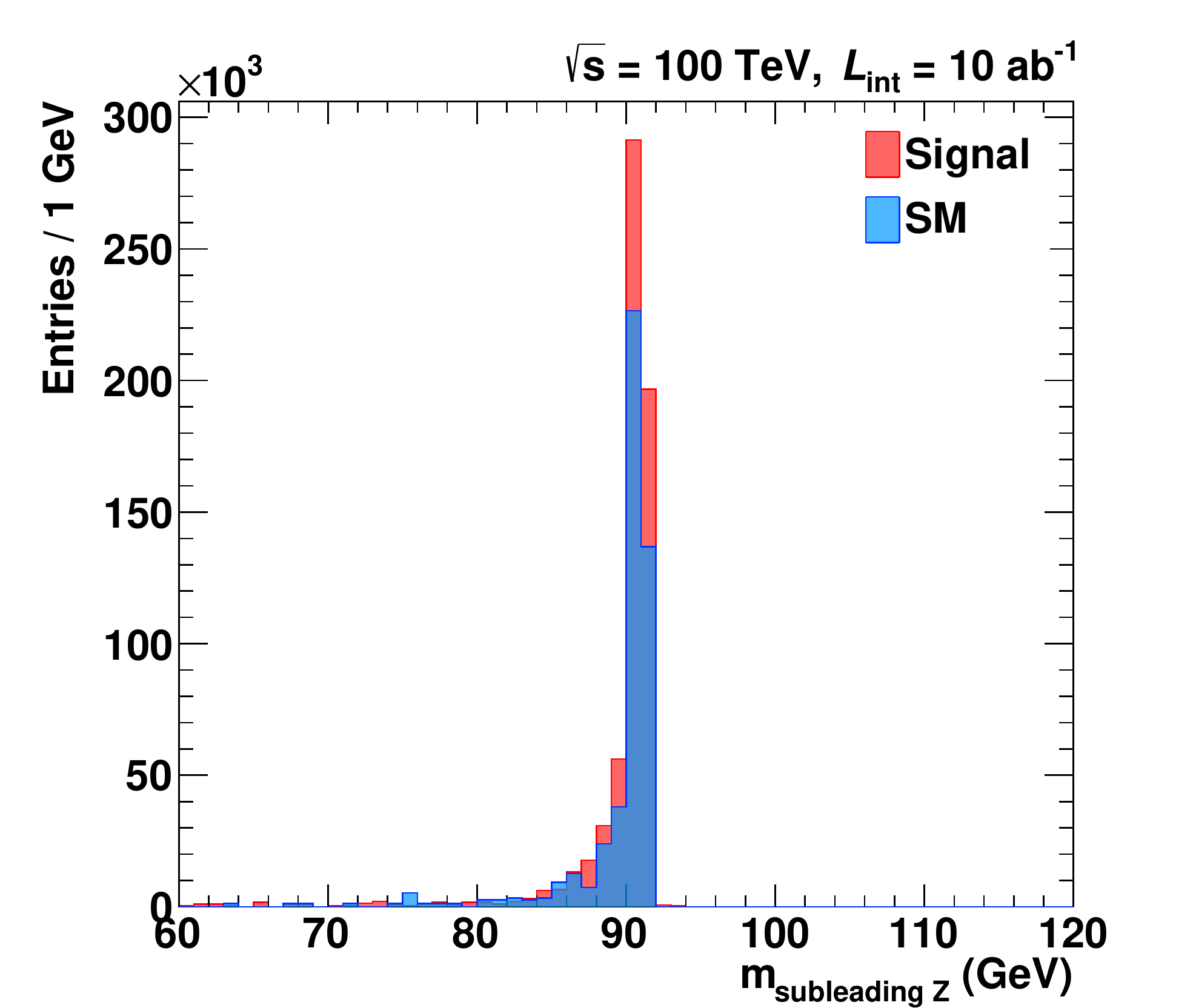}
}
     \caption{Distributions of the reconstructed dilepton candidate mass for four-lepton events selected with both  (a) leading and (b) subleading Z bosons on-shell.} 
   \label{fig:mZ}
\end{figure}
 	
\begin{figure} [hbt]
 \centering
   \includegraphics[width=0.8\textwidth]{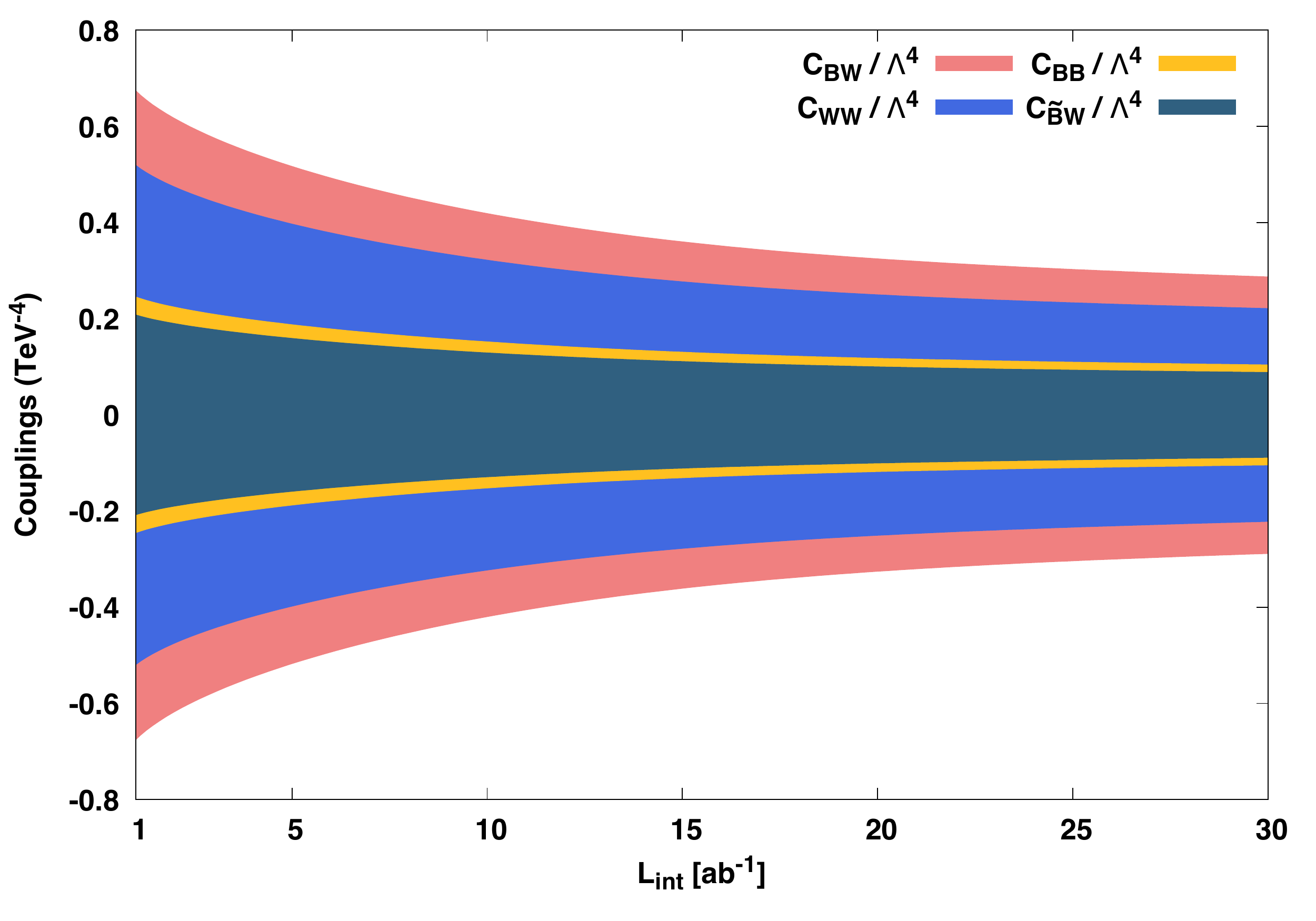}
  \caption{Estimated sensitivity on aNTG couplings at $95\%$ C.L. as a function of integrated luminosity where there is only one coupling varied at a time from its  SM value.} 
   \label{fig:sensitivityLimit}
\end{figure}

For the analysis of $ZZ$ production with $quartet-leptons$ in the final state, the number of signal events and one-parameter $\chi^2$ results for each couplings varied with integrated luminosity from 1 ab$^{-1}$ to 30 ab$^{-1}$. In the analysis, only one coupling at a time is varied from its SM value. 
The results from $\chi^2$ analysis of the couplings describing aTGC interactions of neutral gauge bosons. The coefficients of the operators denoted as $C_{\tilde{B}W}  / \Lambda^{4}$,  $C_{WW} / \Lambda^{4}$, $C_{BW} / \Lambda^{4}$ and  $C_{BB}  /   \Lambda^{4}$ are given in Fig.~\ref{fig:sensitivityLimit}.

We present the results of one-dimensional  95\% C.L. confidence intervals  at $L_{int}$ = 10 ab$^{-1}$  under the assumption that any excess in signal over background  due exclusively to $C_{\tilde{B}W} / \Lambda^{4}$, $C_{WW}  / \Lambda^{4}$, $C_{BW} /  \Lambda^{4}$ or $C_{BB}  /   \Lambda^{4}$ are given in Table~\ref{tab:obtainedLimits}. We also include the effects of systematic errors on the limits.
The obtained limits without systematic errosr are one order better than the current limits  on these couplings of   \texttt{dim-8} operators  converted from the couplings of  \texttt{dim-6} operators for the process $pp \rightarrow ZZ \rightarrow  \ell^+ \ell^- \ell^{\prime +} \ell^{\prime -} $~\cite{PhysRevD.97.032005} at the center of mass energy $\sqrt{s} = 13$ TeV and integrated luminosity $L_{int} = 36.1$ fb$^{-1} $  from the LHC.

\begin{table}[h!]
\centering
\resizebox{\textwidth}{!}{
\begin{tabular}{ lcccc}
 \hline
 Couplings & \multicolumn{4}{c}{Limits at 95\% C.L.} \\
    $ (TeV^{-4})$  & $\delta_{sys}= 0\%$ & $\delta_{sys}= 1\%$ & $\delta_{sys}= 3\%$ & $\delta_{sys}= 5\%$  \\
\hline
$C_{\tilde{B}W}  / \Lambda^{4}$   & $[-0.117, \,\, +0.117]$  & $[-0.315, \,\, +0.315]$  & $[-0.544, \,\, +0.544]$  & $[-0.702, \,\, +0.702]$  \\
$C_{WW}  / \Lambda^{4}$            & $[-0.293, \,\, +0.292]$  & $[-0.805, \,\, +0.805]$ & $[-1.388, \,\, +1.388]$  & $[-1.792, \,\, +1.792]$  \\
$C_{BW} /   \Lambda^{4}$            & $[-0.380, \,\, +0.379]$  & $[-1.036, \,\, +1.036]$ & $[-1.788, \,\, +1.788]$   & $[-2.307, \,\, +2.307]$ \\
$C_{BB}  /   \Lambda^{4}$            & $[-0.138, \,\, +0.138]$  & $[-0.373, \,\, +0.373]$ & $[-0.644, \,\, +0.844]$  & $[-0.831, \,\, +0.831]$  \\
\hline 
\end{tabular} 
}
\caption{Estimated one dimensional 95\% C.L. limits on aNTG couplings with and without a systematic error at $L_{int}$ = 10 ab$^{-1}$. For each single anomalous coupling, all parameters other than the one under study are set to zero.}
\label{tab:obtainedLimits}
\end{table}

\section{Conclusion and Discussion} \label{sec:conclusion}

In this paper we present a phenomenological cut based study for probing the limits on  the CP-conserving $C_{\tilde{B}W} / \Lambda^{4}$ and CP-violating $C_{WW}  / \Lambda^{4}$, $C_{BW} /   \Lambda^{4}$ and $C_{BB} / \Lambda^{4}$  \texttt{dim-8} aNTG couplings via  $ZZ \rightarrow 4\ell$ (where $\ell$ =  $e$ or $\mu$) production at the FCC-hh.

The obtained limits of \texttt{dim-8} aNTG couplings at 95\% C.L. for  $C_{\tilde{B}W} / \Lambda^{4}$, $C_{WW}  / \Lambda^{4}$, $C_{BW} /   \Lambda^{4}$ and $C_{BB} / \Lambda^{4}$ with an $\mathcal{L}_{int}$ = 10 ab$^{-1}$  are one order better than those available prior to this study without systematic error. When we compare these results with the latest search for $\nu \bar{\nu} \gamma$ production~\cite{ATLAS:2018eke} from  the LHC, we have better results on $C_{\tilde{B}W} / \Lambda^{4}$, $C_{WW} / \Lambda^{4}$ couplings and improved results on $C_{BW} / \Lambda^{4}$ and $C_{BB} / \Lambda^{4}$  couplings.



  Even with $5\%$ systematic errors, the obtained bounds for FCC-hh are better than the LHC results on all couplings studied in this paper. The limits of aNTG couplings would benefit from high luminosity and the high energy when the systematic uncertainties are well reduced below $5\%$.

\section{Acknowledgement} \label{sec:acknowledge}

This work was partially supported by Turkish Atomic Energy Authority
(TAEK) under the project grant no. 2018TAEK(CERN)A5.H6.F2-20.

\section*{References}
\bibliography{pp2zzPaper}

\end{document}